%% file: main.tex
\input{ieee_settings}

\input{packages}

\input{watermark}

\begin{document}

\input{title}

\input{authors}

\maketitle

\input{sections/00_abstract}

\input{sections/01_introduction}

\input{sections/02_background}

\input{sections/03_methodology}

\input{sections/04_implementation}

\input{sections/05_results}

\input{sections/06_conclusion}

\input{sections/07_appendix}

\begingroup
\renewcommand{\emph}[1]{\textit{#1}}
\bibliographystyle{IEEEtran}
\bibliography{bibliography/references.bib}
\endgroup

\end{document}

%% file: ieee_settings.tex
\documentclass[journal]{IEEEtran}

%% file: packages.tex
\usepackage[noadjust]{cite}

\usepackage[pdftex]{graphicx}

\usepackage{tikz,pgfplots,pgfplotstable}
\usepackage[dvipsnames]{xcolor}
\usepackage{multirow}
\usepackage{graphicx}
\usepackage{comment}
\usepackage{algorithm}
\usepackage{algpseudocode}
\usepackage{algorithmicx}

\usepackage{listings}
\usepackage{caption}
\usepackage{mathtools}
\usepackage{xfrac}
\usepackage{nicefrac}

\usepackage{bbding}

\usepackage{wasysym}
\usepackage{amssymb}
\usepackage{pifont}

\newcommand{\newcheckmark}{\textrm{\ding{51}}}%

\usepackage{babel}
\usepackage{ulem}
\usepackage{fontawesome}
\usepackage{float}
\usepackage{draftwatermark}

%% file: watermark.tex
\SetWatermarkText{UNDER \\ REVIEW}
\SetWatermarkScale{0.8} 

%% file: title.tex
\title{SAILOR: A Scalable and Energy-Efficient Ultra-Lightweight RISC-V for IoT Security}

\markboth{IEEE Transactions on Computers, Vol. xx, No. x, Month Year}%
{Shell \MakeLowercase{\textit{et al.}}: Bare Demo of IEEEtran.cls for IEEE Journals}

%% file: authors.tex
\author{\IEEEauthorblockN{Christian Ewert, Tim Hardow, Melf Fritsch, Leon Dietrich, Henrik Strunck, Rainer Buchty,\\ Mladen Berekovic, and Saleh Mulhem}\\
\thanks{This work has been partially funded by the Federal Ministry of Research, Technology and Space (BMFTR) via the projects: SASVI (16KIS1577) and RILKOSAN (16KISR010K).}
\thanks{All authors are with the Institute of Computer Engineering, \glqq Universität zu Lübeck\grqq, Germany.}}

%% file: sections/00_abstract.tex
\begin{abstract}
Recently, RISC-V has contributed to the development of IoT devices, requiring architectures that balance energy efficiency, compact area, and integrated security. However, most recent RISC-V cores for IoT prioritize either area footprint or energy efficiency, while adding cryptographic support further compromises compactness. As a result, truly integrated architectures that simultaneously optimize efficiency and security remain largely unexplored, leaving constrained IoT environments vulnerable to performance and security trade-offs.
In this paper, we introduce SAILOR, an energy-efficient and scalable ultra-lightweight RISC-V core family for cryptographic applications in IoT. Our design is modular and spans 1-, 2-, 4-, 8-, 16-, and 32-bit serialized execution data-paths, prioritizing minimal area. This modular design and adaptable data-path minimizes the overhead of integrating RISC-V cryptography extensions, achieving low hardware cost while significantly improving energy efficiency. We validate our design approach through a comprehensive analysis of area, energy, and efficiency trade-offs.
The results surpass state-of-the-art solutions in both performance and energy efficiency by up to 13$\times$ and reduce area by up to 59\,\%, demonstrating that lightweight cryptographic features can be added without prohibitive overhead, and that energy- or area-efficient designs need not compromise performance.
\end{abstract}

%% file: sections/01_introduction.tex
\section{Introduction}

\IEEEPARstart{W}{ith} the rapid evolution of the Internet of Things (IoT), a massive number of battery-powered devices are connected, such as smart cards, a network of sensors, etc. 
This number is expected to surpass $41.1$ billion by 2030~\cite{IoTdevices2030}, leading to new requirements, where energy efficiency and security play a crucial role. 
Here, the energy efficiency indicates new concerns and becomes in high demand when needed for secure IoT communication.
The networked IoT device is required to process, exchange, and transmit a huge amount of sensitive data needed for device identification, authentication, secure communications, and other tasks~(\cite{IoTdevices2030}, \cite{2025_ewert_itrustlet}).

Cryptographic algorithms form the foundation of these security operations and highly impact system utilization and energy consumption. Such security-related computations can consume up to 50\,$\%$ of the total battery energy, thereby constraining device lifetime and network sustainability~(\cite{2011_rifapous_cryptography_energy_costs_handheld_devices, 2025_ewert_itrustlet}).
Consequently, achieving high energy efficiency while maintaining robust security has become a key challenge in IoT processor design \cite{2022_bauer_dualconevmodel}.

To improve the energy efficiency of cryptographic applications, hardware support via cryptography instruction set extensions forms a suitable solution. Such dedicated cryptography instructions enhance the processing capability and energy efficiency for cryptographic workloads while maintaining the flexibility inherent to general-purpose processors.

The RISC-V instruction set was recently extended by dedicated and officially ratified cryptography instruction set extensions for both scalar and vector operations~\cite{riscv_scalar_crypto_v1.0.1,riscv_vector_crypto_v1.0.0}.
The vector-supporting cryptographic extension requires additional vector registers and aims at supporting cryptographic workloads on high-performance CPUs, making it less suitable for resource-constrained and embedded IoT devices. 
In contrast, the RISC-V scalar cryptography extensions are compatible with both 32-bit (RV32) and 64-bit (RV64) ISAs and do not require extra register resources. 

In addition to energy efficiency, stringent area constraints must be addressed when deploying cryptographic applications in resource-constrained IoT environments. For this, integration of lightweight hardware support for cryptographic primitives is essential~(\cite{2020_lightweight_crypto_iot, 2025_najork_singleenginerot}). W.r.t. cryptography instruction set extensions, a compact processor architecture with a low-area-overhead implementation of the cryptography instructions is highly desirable. Therefore, small-area RISC-V cores with low-area-overhead scalar cryptography extensions can be perceived as an excellent opportunity for designing lightweight and energy-efficient processors with security capabilities tailored for battery-powered IoT devices. 

However, existing designs often address either area consumption or energy efficiency in isolation, while cryptographic support tends to be introduced at the expense of compactness.
This determines a research gap that targets a small-area RISC-V core design with lightweight and energy-efficient cryptographic support for resource-constrained IoT devices.
Bridging this gap requires integrated approaches that balance area, energy, and scalar cryptography extensions, thereby enabling the deployment of small and energy-efficient RISC-V processors for secure communication, authentication, and other security-related tasks in highly constrained IoT environments.                                                                               
\subsection{Paper Contribution}
\label{subsec:paper_contribution}

Our paper introduces SAILOR, a new family of RISC-V cores addressing the challenges and provides novel design architectures to bridge the research gap in the current RISC-V design domain. 
Thus, our paper advances the state-of-the-art with a \textbf{small-area and energy-efficient RISC-V architecture with lightweight implementations of cryptography extensions.} Minimizing the area overhead of integrating RISC-V cryptography extensions is very challenging. Here, we employ a bit-serializing approach as a key enabler for a small-area and scalable RISC-V core architecture. Furthermore, we integrate lightweight hardware modules to enhance the performance and energy efficiency for cryptographic workloads. The proposed architecture is scalable and adaptable to the application requirements, balancing out area footprint, energy consumption and performance. Our design methodology presented in Section~\ref{subsec:zkn_zkt_extension} supports RISC-V cryptography extensions, mainly the Zkn-Zkt suite, including Zbkb, Zbkx, Zbkc, Znke, Zknd, and Zknh~\cite{riscv_scalar_crypto_v1.0.1}. 

\textbf{As a proof of concept}, the implementation of the cores in an open-source standard library demonstrates how the proposed architecture achieves a favorable small area footprint, and that integrating lightweight cryptographic extensions does not necessarily impose prohibitive overheads. A state-of-the-art comparison presented in Section~\ref{sec:implementation} provides area trade-off insights and shows security enhancements of the proposed core designs while preserving the compact area demanded by constrained platforms.

\textbf{A comprehensive performance, energy, and efficiency evaluation compared to state-of-the-art RISC-V cores} demonstrates the benefits of the proposed architecture for cryptographic applications when deployed in resource-constrained environments. The analysis in Section~\ref{sec:evaluation} shows significant performance, energy, and efficiency enhancements. This empirical evidence demonstrates the practicality of the proposed approach and its suitability for real-world IoT deployments.

\subsection{Paper Organization}
\label{subsec:paper_organisation}

The paper is structured into four main sections: (I) A foundational section (Section~\ref{sec:background}), providing a technical background, highlighting the state-of-the-art RISC-V core implementations, and showing the motivation and design objectives of our work; (II) Section~\ref{sec:methodology} introduces our design methodology for the proposed family of RISC-V cores, focusing on a lightweight and efficient core architecture, and a low overhead integration of cryptography extensions; (III) Section~\ref{sec:implementation} presents implementation results of our proposed ultra-lightweight RISC-V cores and compares their respective area with state-of-the-art small and cryptography-enhanced cores; (IV) Section~\ref{sec:evaluation} drives quantitative evidence as a proof of concept based on the implementation results of our proposed ultra-lightweight RISC-V cores, where particular focus is on performance, energy expenses, and efficiency. Further insights into the results are given via the tables presented in the Appendix. Section~\ref{sec:conclusion} gives overall conclusions and discusses potential future work.

\input{tables/crypto_extension_overview}

\input{tables/soa_cryptographic_riscv_cores}

%% file: tables/crypto_extension_overview.tex
\begin{table}[t]
\centering
\caption{RISC-V Scalar Cryptography NIST Extensions~\cite{riscv_scalar_crypto_v1.0.1}}
\label{tab:crypto_extension_overview}
\resizebox{\columnwidth}{!}{%
\begin{tabular}{|c|c|}
\hline
\textbf{RISC-V Cryptography Extension Set} & \textbf{Description}                                                                                                                                  \\ \hline
\textit{Zbkb}                          & Bitmanipulation                                                                                                                                           \\ \hline
\textit{Zbkc}                          & Carry-less Multiplication                                                                                                                                 \\ \hline
\textit{Zbkx}                         & Crossbar Permutation                                                                                                                                      \\ \hline
\textit{Zkne} \& \textit{Zknd}         & AES Encryption \& Decryption                                                                                                                              \\ \hline
\textit{Zknh}                          & SHA-256 \& SHA-512                                                                                                                                        \\ \hline
\textit{Zkt}                           & Data Independent Instruction Latency                                                                                                                      \\ \hline
\textit{Zkn}                           & \begin{tabular}[c]{@{}c@{}}NIST Algorithm Suite: \textit{Zbkb}, \textit{Zbkc}, \textit{Zbkx}, \\ \textit{Zkne}, \textit{Zknd}, \textit{Zknh}\end{tabular} \\ \hline
\end{tabular}
}
\end{table}

%% file: tables/soa_cryptographic_riscv_cores.tex
\begin{table*}[t]
\centering
\caption{State-of-the-Art (Lightweight) RISC-V Cores with and without Scalar Cryptography Extensions}
\label{tab:soa_riscv_crypto}
\begin{tabular}{|c|cccccc|cccccc|c|c|}
\hline
\multirow{2}{*}{Reference}                                       & \multicolumn{6}{c|}{NIST Cryptographic Extension (Zkn) Support}                                                                                                                                        & \multicolumn{6}{c|}{Serialization Widths}                                                                                                                                                              & \multirow{2}{*}{\begin{tabular}[c]{@{}c@{}}Energy\\ Efficient\end{tabular}} & \multirow{2}{*}{\begin{tabular}[c]{@{}c@{}}Small\\ Area\end{tabular}} \\ \cline{2-13}
                                                                 & \multicolumn{1}{c|}{Zbkb}          & \multicolumn{1}{c|}{Zbkc}          & \multicolumn{1}{c|}{Zbkx}          & \multicolumn{1}{c|}{Zkne}          & \multicolumn{1}{c|}{Zknd}          & Zknh          & \multicolumn{1}{c|}{1-bit}         & \multicolumn{1}{c|}{2-bit}         & \multicolumn{1}{c|}{4-bit}         & \multicolumn{1}{c|}{8-bit}         & \multicolumn{1}{c|}{16-bit}        & 32-bit        &                                                                             &                                                                       \\ \hline
\cite{serv}                                                      & \multicolumn{1}{c|}{-}             & \multicolumn{1}{c|}{-}             & \multicolumn{1}{c|}{-}             & \multicolumn{1}{c|}{-}             & \multicolumn{1}{c|}{-}             & -             & \multicolumn{1}{c|}{\newcheckmark} & \multicolumn{1}{c|}{-}             & \multicolumn{1}{c|}{-}             & \multicolumn{1}{c|}{-}             & \multicolumn{1}{c|}{-}             & -             & -                                                                           & \newcheckmark                                                         \\ \hline
\cite{qerv}                                                      & \multicolumn{1}{c|}{-}             & \multicolumn{1}{c|}{-}             & \multicolumn{1}{c|}{-}             & \multicolumn{1}{c|}{-}             & \multicolumn{1}{c|}{-}             & -             & \multicolumn{1}{c|}{-}             & \multicolumn{1}{c|}{-}             & \multicolumn{1}{c|}{\newcheckmark} & \multicolumn{1}{c|}{-}             & \multicolumn{1}{c|}{-}             & -             & -                                                                           & \newcheckmark                                                         \\ \hline
\cite{2024_kissich_fazyrv}                                       & \multicolumn{1}{c|}{-}             & \multicolumn{1}{c|}{-}             & \multicolumn{1}{c|}{-}             & \multicolumn{1}{c|}{-}             & \multicolumn{1}{c|}{-}             & -             & \multicolumn{1}{c|}{\newcheckmark} & \multicolumn{1}{c|}{\newcheckmark} & \multicolumn{1}{c|}{\newcheckmark} & \multicolumn{1}{c|}{\newcheckmark} & \multicolumn{1}{c|}{-}             & -             & -                                                                           & \newcheckmark                                                         \\ \hline
\cite{picorv32}                                                  & \multicolumn{1}{c|}{-}             & \multicolumn{1}{c|}{-}             & \multicolumn{1}{c|}{-}             & \multicolumn{1}{c|}{-}             & \multicolumn{1}{c|}{-}             & -             & \multicolumn{1}{c|}{-}             & \multicolumn{1}{c|}{-}             & \multicolumn{1}{c|}{-}             & \multicolumn{1}{c|}{-}             & \multicolumn{1}{c|}{-}             & \newcheckmark & -                                                                           & \newcheckmark                                                         \\ \hline
\cite{2020_marshall_implementing_draft_crypto}                   & \multicolumn{1}{c|}{$\circ$}       & \multicolumn{1}{c|}{$\circ$}       & \multicolumn{1}{c|}{$\circ$}       & \multicolumn{1}{c|}{$\circ$}       & \multicolumn{1}{c|}{$\circ$}       & $\circ$       & \multicolumn{1}{c|}{-}             & \multicolumn{1}{c|}{-}             & \multicolumn{1}{c|}{-}             & \multicolumn{1}{c|}{-}             & \multicolumn{1}{c|}{-}             & \newcheckmark & -                                                                           & -                                                                     \\ \hline
\cite{2023_gewehr_improving_efficiency_cryptographic_algorithms} & \multicolumn{1}{c|}{-}             & \multicolumn{1}{c|}{-}             & \multicolumn{1}{c|}{-}             & \multicolumn{1}{c|}{\newcheckmark} & \multicolumn{1}{c|}{-}             & \newcheckmark & \multicolumn{1}{c|}{-}             & \multicolumn{1}{c|}{-}             & \multicolumn{1}{c|}{-}             & \multicolumn{1}{c|}{-}             & \multicolumn{1}{c|}{-}             & \newcheckmark & \newcheckmark                                                               & -                                                                     \\ \hline
\cite{2022_nisanci_symmetric_crypto_riscv}                       & \multicolumn{1}{c|}{$\circ$}       & \multicolumn{1}{c|}{\newcheckmark} & \multicolumn{1}{c|}{\newcheckmark} & \multicolumn{1}{c|}{\newcheckmark} & \multicolumn{1}{c|}{\newcheckmark} & \newcheckmark & \multicolumn{1}{c|}{-}             & \multicolumn{1}{c|}{-}             & \multicolumn{1}{c|}{-}             & \multicolumn{1}{c|}{-}             & \multicolumn{1}{c|}{-}             & \newcheckmark & -                                                                           & -                                                                     \\ \hline
{[}This work{]}                                                  & \multicolumn{1}{c|}{\newcheckmark} & \multicolumn{1}{c|}{\newcheckmark} & \multicolumn{1}{c|}{\newcheckmark} & \multicolumn{1}{c|}{\newcheckmark} & \multicolumn{1}{c|}{\newcheckmark} & \newcheckmark & \multicolumn{1}{c|}{\newcheckmark} & \multicolumn{1}{c|}{\newcheckmark} & \multicolumn{1}{c|}{\newcheckmark} & \multicolumn{1}{c|}{\newcheckmark} & \multicolumn{1}{c|}{\newcheckmark} & \newcheckmark & \newcheckmark                                                               & \newcheckmark                                                         \\ \hline
\end{tabular}
\label{tab:soa_riscv_crypto}

\footnotesize{-: the RISC-V core does \textbf{not} cover this feature. $\circ$: this feature is partially implemented, for example, a draft version of\\ the cryptography extensions in~\cite{2020_marshall_implementing_draft_crypto} was implemented and not the ratified one. \newcheckmark: the feature is implemented.}
\end{table*}

%% file: sections/02_background.tex
\section{Background, Related Work \& Motivation}
\label{sec:background}

This section reviews RISC‑V cryptographic extensions and shows state‑of‑the‑art implementations. Then, serialized data processing is presented as an approach to minimize the area consumption for lightweight core designs. Finally, our motivation and design objectives for the proposed cores are established.

\subsection{RISC-V Scalar Cryptography Extensions}
\label{subsec:riscv_crypto_extension}

To improve performance and efficiency of cryptographic applications, the RISC-V ISA has recently been extended with dedicated cryptographic extensions for both scalar and vector operations, which have been officially ratified for standard use~\cite{riscv_scalar_crypto_v1.0.1,riscv_vector_crypto_v1.0.0}. The RISC-V scalar cryptography extensions are compatible with both 32-bit (RV32) and 64-bit (RV64) ISAs, and incorporate the NIST Algorithm Suite \textit{Zkn}~\cite{riscv_scalar_crypto_v1.0.1}, which includes bit‑ and byte‑oriented operations, Advanced Encryption Standard (AES), and Secure Hash Algorithm 2 (SHA-2), spanning a wide range of cryptographic applications. 
As detailed in Table~\ref{tab:crypto_extension_overview}, key operational subsets include bit manipulation (\textit{Zbkb}), crossbar permutation (\textit{Zbkx}), and carry-less multiplication (\textit{Zbkc}). 
It also includes dedicated instructions for AES encryption and decryption (\textit{Zkne} and \textit{Zknd}), and SHA-2 hash computation (\textit{Zknh}).
To mitigate timing side-channel vulnerabilities inherent to data‑dependent instruction execution times, the Data-Independent Execution Latency Subset \textit{Zkt} mandates constant-time implementation for pertinent instructions, requiring execution latency to be independent of the processed data. These architectural features enable, particularly when applied to the RV32I ISA~\cite{riscv_rv32i_v2.1}, the design of compact, low-power RISC-V cores that are highly suitable for secure and efficient IoT deployments. 

State-of-the-art RISC-V processor designs emphasize performance enhancements for specific cryptographic primitives, such as AES, cryptographic hash functions, and digital signature generation. Acceleration relies mainly on memory-mapped IPs~(\cite{2024_simola_riscv_aes256_acc,2025_van_tinh_aes_rv, 2024_duc_hong_crypto_coprocessor_sec_app, 2024_vu_trung_rvcp}) or customized loosely-coupled co-processors~(\cite{2021_zgheib_aes_hw_acc_iot, 2021_wang_energy_efficient_crypto_extension}), often without addressing resource‑constrained devices and their area requirements. 
Other approaches propose customized extensions, either dedicated for AES~\cite{2020_marshall_scalar_aes_extention_ise, 2024_ignatius_power_side_channel_attack_crypto_core} or jointly supporting AES and SM4~\cite{2020_saarinen_lightweight_isa_extension_aes}, deeply integrating functionality of cryptographic primitives into RISC-V cores. Just a few works provide a holistic perspective and consider the implementation of a complete cryptographic extension.
In \cite{2020_marshall_implementing_draft_crypto}, a draft version of the RISC-V cryptographic extension within the SCARV processor~\cite{scarv} is implemented. The results show a 17\,\% area overhead and up to 3$\times$ speedup of cryptographic algorithms, but without insights into energy enhancements. Similarly, the results presented in~\cite{2022_nisanci_symmetric_crypto_riscv} show a 3.6$\times$ performance gain for AES at the cost of 37.8\,\% hardware overhead, again without energy analysis.
The approach presented in~\cite{2023_gewehr_improving_efficiency_cryptographic_algorithms} achieves performance improvements of 44$\times$ and energy reductions of 28$\times$ for AES encryption by integrating crypto extensions into the Ibex processor~\cite{ibex}. However, it does not consider hardware support for AES decryption. These approaches focus on enhancing performance and energy while overlooking area requirements, which is highly desirable for resource-constrained devices. This shows open challenges and highlights the need for lightweight RISC-V processors specifically tailored for cryptographic applications.

\subsection{RISC-V Bit-Serialized Data Path Architecture}
\label{subsec:serialized_data_processing}

The current majority of RISC-V processors adhere to the RV32I ISA by utilizing an internal 32-bit data path. However, serialization approaches were recently proposed for area minimization by shrinking the internal data path, i.e., the data is processed in smaller chunks. The serialization approach incorporates calculating instruction results in multiple cycles, where the size of the arithmetic-logical unit (ALU) and functional modules is adjustable according to the operand width. For an efficient computation, the data-width is normally divided into equal-sized chunks, e.g., 1-, 2-, 4-, and 8-bit. A prominent example is the award-winning SERV~\cite{serv} employing a 1-bit wide data path while retaining RV32I ISA compatibility. This approach for area minimization has lead to map over 4000 cores on an FPGA~\cite{plenticore_serv}. This architecture has been enhanced for better performance to the QERV processor~\cite{qerv}, which extends the serialized data path to 4 bits.
These approaches are highly optimized for area minimization, each committed to a fixed serialization width.

For enhanced design flexibility, FazyRV~\cite{2024_kissich_fazyrv} introduces a scalable and parameterizable architecture capable of supporting 1-, 2-, 4-, and 8-bit wide data paths, thereby addressing a broader range of serialized implementations. They focus on minimizing area consumption, especially when deployed on FPGAs, and do not emphasize optimized ASIC designs. However, deploying processors for IoT applications, not only requires a small area but also desires minimal energy consumption. To maintain performance, furthermore, area and energy efficiency must be taken into account. 

\subsection{Motivation \& Design Objectives}
\label{subsec:motivation}

The performed review and analysis of the state-of-the-art RISC-V core designs and addressing their challenges are summarized in Table~\ref{tab:soa_riscv_crypto}.
This leads to the following key observations: 

\begin{itemize}
    \item \textit{Observation 1}: Few RISC-V core implementations incorporate comprehensive cryptography extensions (e.g.,~\cite{2020_marshall_implementing_draft_crypto, 2023_gewehr_improving_efficiency_cryptographic_algorithms, 2022_nisanci_symmetric_crypto_riscv}). These works emphasize performance and energy-efficiency improvements for RISC-V 32-bit cores, positioning them as principal design objectives. However, such implementations frequently overlook area requirements, despite their critical role in determining deployment for resource-constrained IoT devices.

    \item \textit{Observation 2}: Existing small-area RISC-V core implementations (e.g.,~\cite{serv,qerv,2024_kissich_fazyrv,picorv32}) predominantly prioritize area minimization, while overlooking energy requirements. Compact core architectures are essential for the integration of processors into resource-constrained environments. Anyway, IoT platforms extend beyond achieving a small area; stringent energy requirements must also be addressed, as energy efficiency directly impacts device autonomy and long-term deployment. Therefore, future design efforts should consider both energy and area efficiency as co-primary objectives for the deployment of cryptographic applications on IoT devices.
\end{itemize}

These observations indicate a research gap that motivates our work to target a small-area and energy- and area-efficient core implementation supporting the deployment of cryptographic applications on IoT devices. The design of the proposed scalable SAILOR cores covers 1-, 2-, 4-, 8-, 16-, and 32-bit widths as (serialized) data paths utilizing the RISC-V RV32I ISA. We enhance the performance and energy consumption for cryptographic workloads by integrating the RISC-V cryptography Zkn and Zkt extensions w.r.t. minimal hardware overhead and a prevention of timing side-channel vulnerabilities.
In detail, we aim at the following design objectives (\textit{Obj}):

\begin{figure*}[t]
  \centering
  \includegraphics[width=0.9\textwidth]{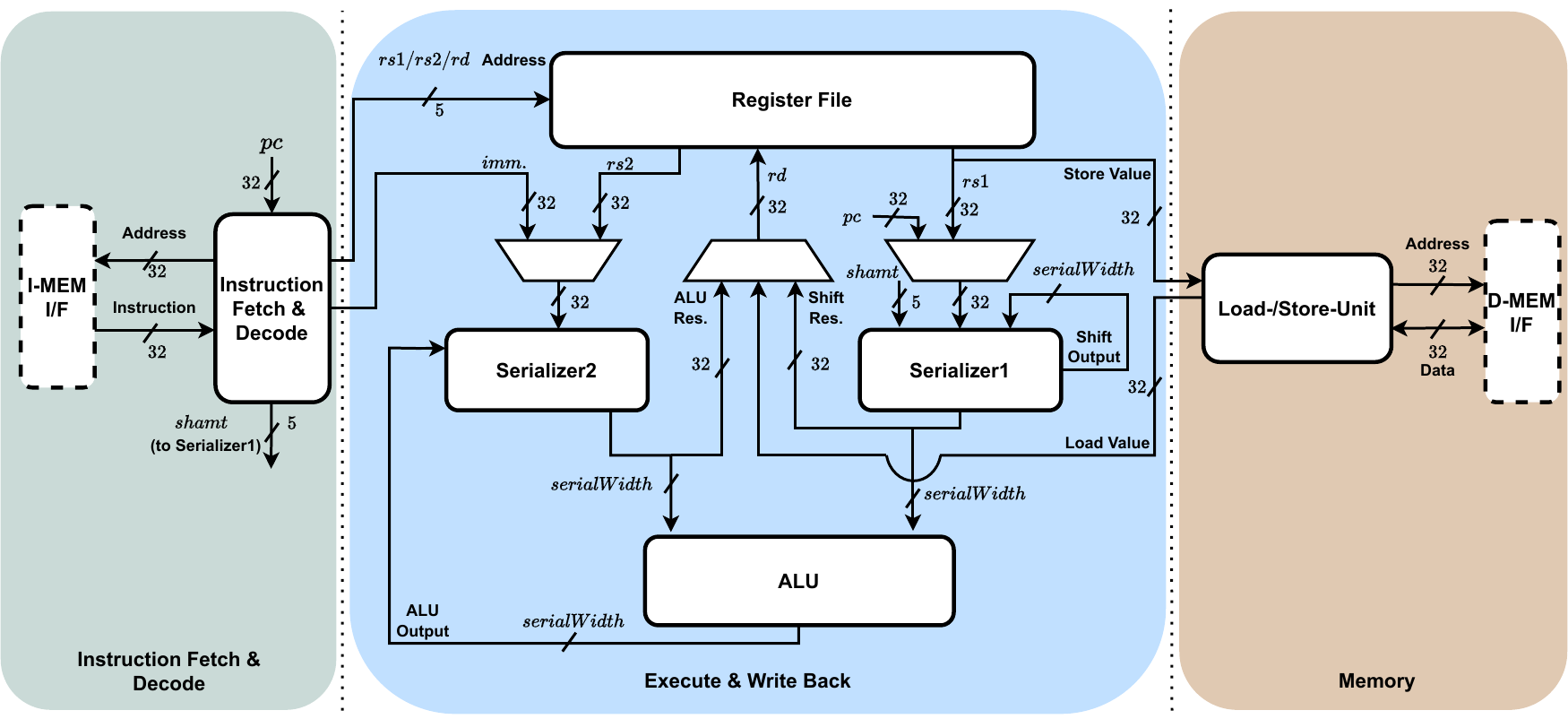}
  \caption{Architecture of the RV32I serialized SAILOR}
  \label{fig:architecture_serialized_rv32i_cpu}
\end{figure*}

\subsubsection*{Obj. 1) Area- \& Energy-Efficiency}
\label{subsubsec:area_energy_efficiency}

Unlike the state-of-the-art RISC-V core implementations~\cite{2020_marshall_implementing_draft_crypto, 2023_gewehr_improving_efficiency_cryptographic_algorithms, 2022_nisanci_symmetric_crypto_riscv, serv, qerv, 2024_kissich_fazyrv, picorv32},  the proposed architecture should be explicitly optimized w.r.t. both area- and energy-efficiency. This can be perceived as a design strategy enabling a balanced trade-off between area footprint and energy consumption.

\subsubsection*{Obj. 2) Scalability \& Adaptability}
\label{subsubsec:configurability_modularity}

A RISC-V core design approach that can be easily adaptable to the requirements of desired cryptographic IoT applications. The serialized data path allows scalability, while a modular design approach can facilitate the seamless integration of cryptography extensions into the baseline RV32I architecture. This allows the processor to be tailored to the specific cryptographic application requirements.

\subsubsection*{Obj. 3) Performance}
\label{subsubsec:performance}

The design can take advantage of several design routines to improve performance, such as an instruction fetch buffer, dedicated ISA extensions for cryptographic workloads, or reusing existing functional units. These further contribute to the overall efficiency as long as the enhancements predominate the introduced area overhead.

%% file: sections/03_methodology.tex
\section{Design Methodology}
\label{sec:methodology}

\input{algorithms/serial_data_path_description}

This section introduces the proposed modular and scalable design of the RISC‑V cores called SAILOR, covering the RV32I ISA and its Zkn–Zkt cryptography extension. 
Due to its inherent modularity, the proposed architecture enables independent scaling of the data path and instantiation of cryptographic hardware support.
This flexibility supports efficient cryptographic acceleration and fulfills the requirements of secure and resource-constrained IoT devices.

\subsection{Hardware Design of the RV32I SAILOR Cores}
\label{subsec:rv32i_baseline_architecture}

The RV32I baseline core architecture, illustrated in Fig.~\ref{fig:architecture_serialized_rv32i_cpu}, can be partitioned into three primary stages: (i) the instruction fetch and decode stage, (ii) the data memory stage including a dedicated load/store unit, and (iii) the (serialized) execution and write-back stage.

To ensure high modularity, the memory and register file interface data widths are tailored towards 32-bit and not serialized. The abstraction of the memory interfaces ensures compatibility with a variety of system buses, such as Tilelink, AXI4, or Wishbone, without requiring structural modifications to the core. Similarly, the detachment of the register file from the (serialized) data path simplifies integration of architecture-independent register files. Thus, the architecture facilitates targeted optimizations within the data path itself. The modularity provides flexibility for extending the baseline implementation with additional functional units required for ISA extensions, while preserving interface consistency and minimizing integration overhead.

\subsubsection{Serialized Data Processing}
\label{subsubsec:serial_data_processing}

The serialized data processing path is described in Alg.~\ref{alg:serial_data_path}. It composes three fundamental modules: \textit{Serializer1}, \textit{Serializer2}, and the arithmetic–logic unit (ALU). The Serializer1 and Serializer2 divide the input data into data chunks of width defined by the configuration parameter \textit{serialWidth}. The ALU performs the chunk-wise computations on the serialized data. For register–register operations, the 32-bit operands \textit{rs1} and \textit{rs2} are loaded into Serializer1 and Serializer2, respectively. In the case of register–immediate operations, the 32-bit extended immediate value encoded in the instruction is loaded into Serializer2. Once both Serializer1 and Serializer2 are initialized, the computation proceeds sequentially on data chunks, starting with the least significant serialWidth-bits (LSBs). At each computation step, the remaining data chunks are shifted toward the LSB, while Serializer2 preserves the partial result. When all data chunks are processed, the instruction result is available in Serializer2 and written back to the register file. The ALU architecture is adapted to the serialWidth data-path, such that it can handle one serialized data chunk within a single clock cycle.

Serializer1 not only provides the chunk-wise LSB-directed shifts required to supply data to the ALU. It is also enhanced to support shift instructions natively. For this, Serializer1 is extended with single-bit shift functionality, enabling correct execution when the shift amount (shamt) cannot be fully realized by chunk‑wise data shifting alone. This mechanism natively supports right-shift operations. To increase both efficiency and performance, Serializer1 is further extended with left-shift functionality, shifting the data towards the most significant bit (MSB). Consequently, Serializer1 functions as a bidirectional shift register. In practice, this extension reduces the execution latency of left-shift instructions by up to 50\% and eliminates the overhead of preserving intermediate shifted values in auxiliary registers.

The modular design of the architecture also facilitates the implementation of a fully parallel 32-bit data path. In this configuration, operands are supplied directly from the register file to the ALU, and the computation result is written back to the register file within a single clock cycle. However, Serializer1 is retained for shift operations. The division of shifts into larger steps and single-bit shifts remains configurable, allowing an adjustment of the trade-off between area consumption, performance, and efficiency.

\subsubsection{Instruction Fetch \& Branch Prediction}
\label{subsubsec:instruction_decoding}

We enhance the architecture's performance and efficiency by adding a small fetch buffer holding the succeeding instruction. When the current instruction execution is accomplished, control signals from the prefetched instruction, and operands from the register file are loaded, while the result of the previous instruction is written back to the register file.
This increases the efficiency of proposed cores compared to the state-of-the-art serialized cores~(\cite{serv,qerv,2024_kissich_fazyrv}), as their main focus is only on a small area.
To enable instruction fetching, the program counter ($pc$) is updated via a dedicated branch unit. For a lightweight realization with low area overhead, a forward branch prediction under the assumption of word-aligned instructions is deployed. It should be noted that our modular architecture allows us to integrate more sophisticated branch-prediction mechanisms if required. A comprehensive evaluation of their impact on area, performance, energy consumption, and overall efficiency is beyond the scope of this work and, therefore, omitted.  

\subsection{Hardware Design for the Zkn-Zkt Extension}
\label{subsec:zkn_zkt_extension}

\begin{figure*}[t]
  \centering
  \includegraphics[width=0.9\textwidth]{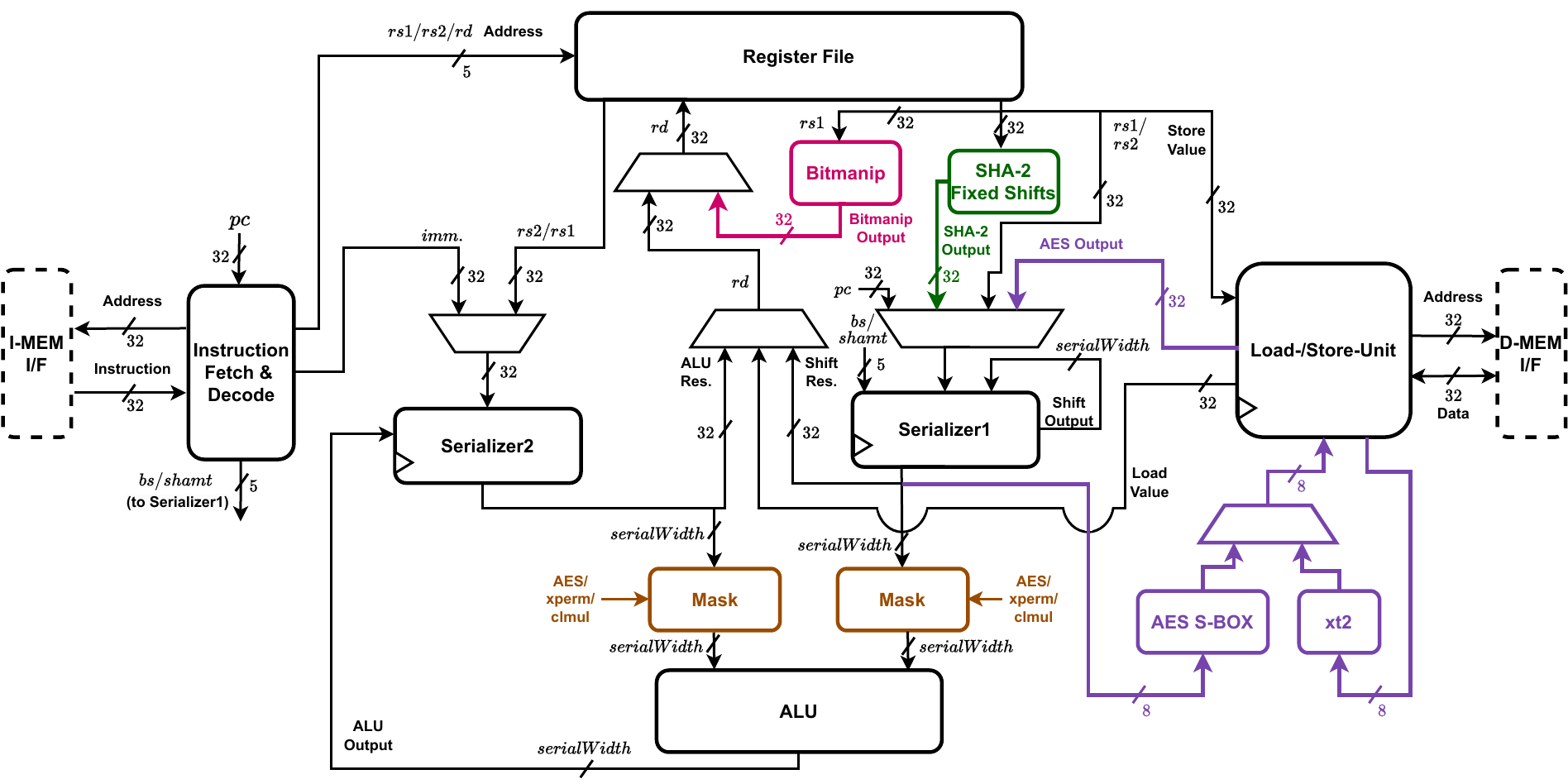}
  \caption{Architecture of the RV32I-Zkn-Zkt serialized SAILOR}
  \label{fig:architecture_serialized_rv32i_crypto_cpu}
\end{figure*}

We extend the proposed SAILOR cores with RISC-V cryptography instruction set extensions \textit{Zkn-Zkt}. 
Our Zkn-Zkt implementation aims at minimal hardware overhead, enhancing energy consumption and efficiency for secure IoT applications. Due to our modular design approach, all partial extensions can be enabled independently during design time, thereby adapting the core to meet the application's requirements. The overall architecture with all cryptography extensions enabled is illustrated in Fig.~\ref{fig:architecture_serialized_rv32i_crypto_cpu}. 

Following our design objectives, the Zkn-Zkt implementation relies on reusing existing functionality of the baseline RV32I architecture while incorporating lightweight hardware extensions to improve performance, energy consumption, and efficiency. In particular, the ALU provides basic logical functions that are leveraged within cryptographic primitives. 
When possible, existing modules such as the ALU and the Serializer are extended to enhance cryptographic functionality. 
Lightweight auxiliary modules with minimal area overhead are integrated to support cryptographic operations. 
The architecture includes dedicated modules for the AES extensions Zkne \& Zknd, SHA-2 extension Zknh, and bit-manipulation extension Zbkb. 

\begin{figure}[b]
    \centering
    \includegraphics[width=\columnwidth]{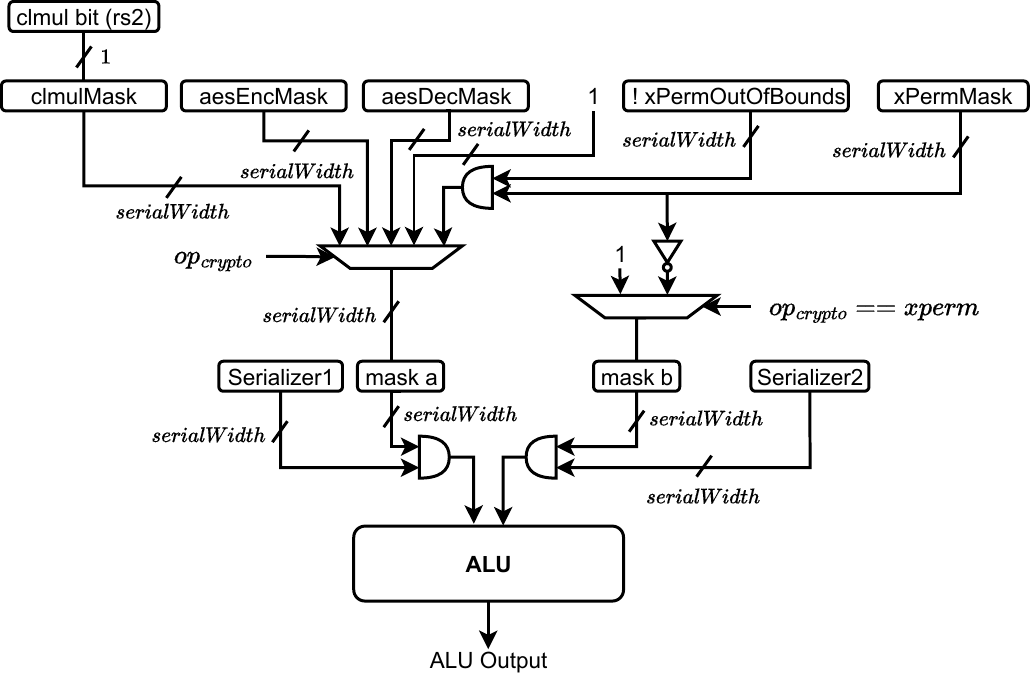}
    \caption{Combined ALU Mask for Cryptographic Extensions}
    \label{fig:alu_with_mask}
\end{figure}

A key operational module is the ALU operand mask as shown in Fig.~\ref{fig:alu_with_mask}. It selects appropriate data chunks from the serializer to be processed in the ALU. The mask functionality is particularly relevant for carry-less multiplication (cmul/cmulh) and crossbar permutation (xperm) instructions. In the case of cmul/cmulh, the multiplication bit of the rs2 multiplier determines if the shifted rs1 multiplicand is accumulated to the result buffered in Serializer2. In the case of xperm instructions, the mask controls which nibble or byte of rs1, as indexed by rs2, is stored in Serializer2. The realization of cryptography extensions deploys the mask mechanism and others, as outlined in the following:

\subsubsection{Hardware Design for the Zkne \& Zknd Extension}
\label{subsubsec:zknd_zkne_extension}

\begin{figure}[bt]
    \centering
    \includegraphics[width=\columnwidth]{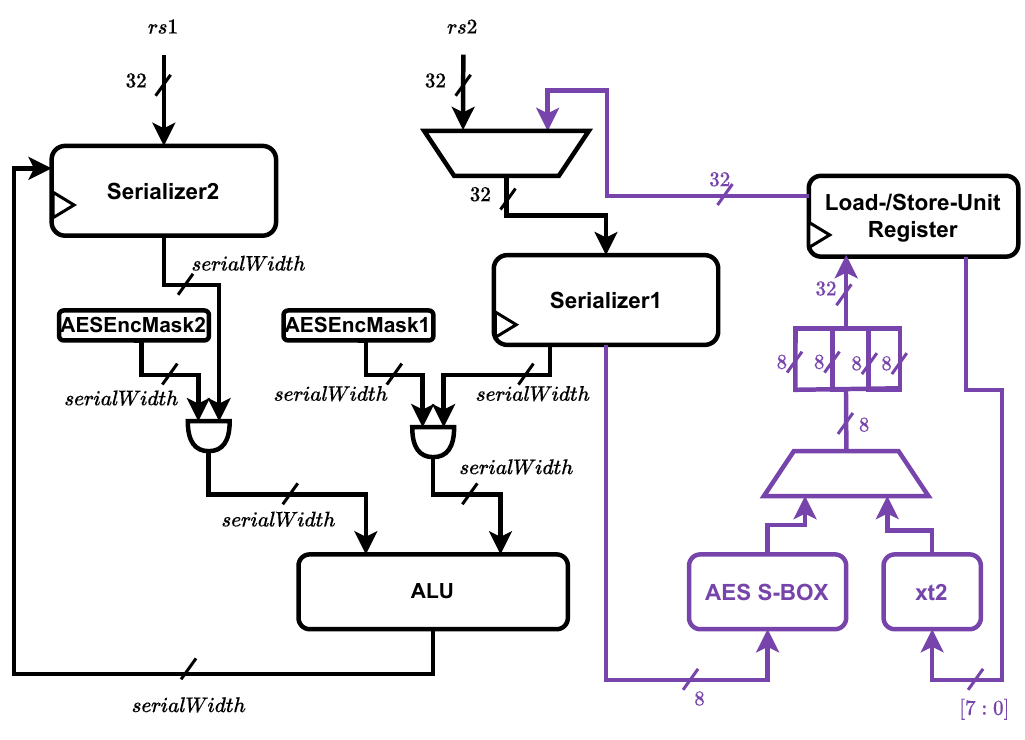}
    \caption{The AES Encryption Extension Data Path}
    \label{fig:aes_zkne_datapath}
\end{figure}

AES is a NIST-standardized 10-round symmetric block cipher with different key lengths of 128, 192, or 256 bits and a 128-bit fixed data block~\cite{aes_nist_standard}.
The AES encryption and decryption extension (Zkne/Zknd) comprises operations supporting middle round encryption (\textit{aes32esmi}) and decryption (\textit{aes32dsmi}), as well as the corresponding final round instructions (\textit{aes32esi}/\textit{aes32dsi})~\cite{riscv_scalar_crypto_v1.0.1}.
Implementing the AES extensions requires two additional hardware units as illustrated in Fig.~\ref{fig:aes_zkne_datapath}. 

First, an AES S-Box is integrated as a lightweight hardware extension. 
Here, the state-of-the-art smallest combined S-Box for encryption and decryption~\cite{2019_maximov_circuit_minimzation_smaller_faster_sboxes} is implemented to ensure minimal hardware overhead. The lightweight dedicated S-Box module achieves single-cycle execution at minimal hardware cost, thus reducing latency, energy consumption, and increasing efficiency.
Second, a \textit{xtime} unit (xt2) efficiently performs multiplications over the Galois field. 
Both units reuse the buffer register present in the load-/store-unit, waiving the need for auxiliary registers required to perform round encryption and decryption operations.
Furthermore, efficient byte selection and word rotation are performed by the ALU operand mask in conjunction with Serializer1, which is enhanced with lightweight rotation operation support. This enhancement also involves bit-manipulation instructions from the Zbkb extension set.

By leveraging the proposed modular design approach, adding lightweight hardware modules, and reusing existing functionality of the baseline architecture, the implementation aims at a trade-off between maintaining low area and efficiently enhancing performance and energy consumption. 

\subsubsection{Hardware Design for the Zknh \& Zbkb Extension}
\label{subsubsec:zknh_extension}

Secure Hash Algorithm 2 (SHA-2) is a family of cryptographic hash functions standardized by NIST to ensure the integrity of transmitted data. 
SHA-2 has two variants: SHA-256 and SHA-512, indicating the bit size of output values (digests)~\cite{sha2_nist_standard}.
To realize SHA-2, hash and bit-manipulation instructions are incorporated. 
A particular focus is on fixed shifts and permutation operations required for SHA-2.
Instead of reusing the generic shift and rotation logic of the Serializer, the design employs specialized fixed shifts and rotations that complete within a single clock cycle.  
Reusing the existing ALU functionality, the XOR operation can be implemented directly via the ALU, thereby limiting the introduced hardware overhead.
Similarly, (re-)ordering instructions such as \textit{(un-)zip} and \textit{(b)rev8} exploit single‑cycle implementations, since their mapping positions are fixed. 
The operand can therefore be processed directly and written back to the register file, without passing through the Serializer. This approach enables efficient realization of these instructions while incurring only minimal hardware overhead.

\subsubsection{Hardware Support for the Zkt Extension}
\label{subsubsec:zkt_extension}

To satisfy the constant‑time execution requirements of the Zkt extension, the entire data bytes and words should always be processed at once in the case of AES and SHA-2, respectively. Algorithmic, logical, shift/rotation, and permutation operations exhibit identical latency, regardless of operand values or shift amounts.
This design strategy mitigates core architecture‑related timing side channels with negligible countermeasure overhead.

\subsection{CSR \& Interrupt Support}
\label{subsec:csr_interrupt_support}

Our modular design approach allows including control status registers (CSRs) and interrupt capabilities. We provide full external interrupt and exception support and attach a CSR module that includes all machine-mode CSR specified in the \textit{Zicsr} extension~\cite{riscv_priv}. These extensions allow for full system integration and can be included independently if required.

%% file: algorithms/serial_data_path_description.tex
\begin{algorithm*}[t]
\begin{algorithmic}[1]
\Procedure{Serial Data Processing}{$rd,~rs1,~rs2,~op,~serialWidth$}
\State $s \gets serialWidth$
\State Serializer1[31:0] $\gets rs1[31:0]$ \Comment{Load Operands from the Register File into the Serializer}
\State Serializer2[31:0] $\gets rs2[31:0]$
\For{$i \gets \nicefrac{32}{s}-1$ downto $0$} \Comment{Process Data Chunks}
    \State ALU$~\gets~op$, Serializer1[$s-1:0$], Serializer2[$s-1:0$] \Comment{Process Data Chunks by the ALU}
    \For{$j \gets \nicefrac{32}{s}-1$ downto $1$}
        	\State $k \gets j\times s$
        	\State Serializer1[$(k-1):(k-s)$] $\gets$ Serializer1[$(k+s-1):k$] \Comment{Shift Down Data Chunks}
        	\State Serializer2[$(k-1):(k-s)$] $\gets$ Serializer2[$(k+s-1):k$]
    \EndFor
    \State Serializer2[$31:(32-s)$] $\gets$ ALU \Comment{Preserve Data Chunk Result in Serializer2}
\EndFor
\State $rd[31:0] \gets$ Serializer2[31:0] \Comment{Write Instruction Result in the Register File}

\EndProcedure
\end{algorithmic}
\caption{Description of the Serialized Data Path}
\label{alg:serial_data_path}
\end{algorithm*}

%% file: sections/04_implementation.tex
\section{Implementation \& Area-Evaluation}
\label{sec:implementation}

In this section, the implementation results of the proposed SAILOR cores and area comparisons with the state-of-the-art are presented. 
We implement the proposed scalable and modular RISC-V core architecture using \textit{Chisel} hardware-construction language \cite{chisel_book}. \textit{Chisel} enables the configurability, scalability, and modularity of our architecture while preserving the maintainability of the code base, due to its higher-level abstractions and software-engineering-oriented concepts. Leveraging Chisel’s object-oriented features and parametric class structure, architectural modules are parameterized and extended efficiently. In the RV32I configuration, the ALU supports all basic arithmetic and logical operations. When Zkn extensions are included, an enhanced ALU implementation is derived through inheritance, with configuration parameters passed from the top-level design to the ALU module. This approach preserves the baseline architecture while enabling the modular integration of additional functionality without introducing modifications to the core design.

\subsection{Verification}
\label{subsec:verification}

Dynamic verification of the SAILOR cores is carried out using the \textit{RISC-V Torture Test} framework~\cite{riscv_torture} targeting both RTL and gate-level netlist representations. During simulation, the CPU computes the verification signature, which is then compared with a reference signature generated by the Spike RISC-V simulator~\cite{riscv_sim_spike}. To ensure coverage of cryptography extensions, the torture test generator was expanded to include corresponding Zkn instructions, enabling dynamic verification of the cryptography instruction set extended cores.

\subsection{Area Analysis of the SAILOR Cores}
\label{subsec:area_analysis_sailor}

We evaluate and compare the synthesized gate-level netlist area of serialized core configurations with 1-, 2-, 4-, 8-, and 16-bit data-paths, alongside the full 32-bit data-path implementation, each configured with the corresponding cryptography extensions. In the 32-bit data-path design, \textit{Serializer1} supports both 8-bit serialized and single-bit shifts. \textit{Serializer2} is omitted since additional serialization is unnecessary. Control and status registers (CSRs) are not included in this study, as their optimization is out of the scope of this paper.

\input{diagrams/area_sailchip}

For a detailed area analysis, we synthesized the proposed SAILOR cores with the Synopsys Design Compiler 2024.09-SP2~\cite{synopsys_design_compiler} using the FreePDK45\textsuperscript{TM} 45\,nm open-cell library~\cite{freepdk45} at a typical corner-case temperature of 25$^{\circ}$C, voltage of 1.1\,V, and 100\,MHz operating frequency. The corresponding results are presented as gate equivalence (GE) in Fig.~\ref{fig:area_comparison_endavour}. Serialization introduces an additional area overhead in the range of 1.10\,\% to 2.55\,\%, with the 1-bit serialized design yielding the smallest implementation. In contrast, the full 32-bit data-path achieves a 2.32\,\% area reduction relative to the 1-bit serialized core. These observations highlight the trade-off between data-path width and area efficiency, indicating that wider data-paths may compensate for serialization overhead when auxiliary components are omitted. 

We investigate the hardware overhead associated with the integration of the dedicated cryptographic ISA extensions. In the serialized core configurations, these extensions increase the area by 1.3\,\% to 10.14\,\%. Among them, the \textit{Zbkx} extension exhibits the lowest overhead, as it solely requires the inclusion of the ALU mask unit for byte- and nibble-selection. The largest overhead arises from the \textit{Zknh} extension, where the multiplexer required to select between fixed shift and rotation results constitutes the dominant costs. When the complete \textit{Zkn–Zkt} configuration is enabled, the area overhead increases by up to 21.9\,\% for the serialized 1- to 16-bit and 38.8\,\% for the 32-bit core. Here, the 32-bit core occupies 8.13\,\% more area than the largest serialized design. In this configuration, dedicated adaptations are required to integrate additional modules. For example, \textit{Serializer1} is employed not only for shift operations but also for result accumulation. Such an adaptation, while intrinsic to the serialized designs, must be explicitly integrated into the 32-bit data-path implementation.

The results demonstrate the advantages of a serialized architecture, which inherently facilitates the integration of hardware support for cryptographic applications. The 32-bit data path architecture incorporates additional resources for the integration of cryptography extensions, which consequently results in increased area overhead as compared to the serialized approaches. The configurable and adaptable design approach allows core configurations to be tailored to specific applications while maintaining a compact area and selectively incorporating extensions to enhance performance and efficiency. 

\subsection{Area Comparison with State-of-the-Art RV32I Cores}
\label{subsec:area_comparison_soa_rv32i}

\input{tables/area_comparison_rv32i}

Serialized implementations of the RISC‑V architecture have recently been investigated as promising candidates for low‑area core designs \cite{2024_kissich_fazyrv, serv, qerv}. In this work, we evaluate the proposed SAILOR RV32I cores with serialized data-paths of 1-, 2-, 4-, and 8-bit, along with their respective cryptography extensions \textit{Zkn–Zkt}. These designs are compared against state‑of‑the‑art serialized architectures, including the 1‑bit SERV~\cite{serv}, its optimized 4‑bit variant QERV~\cite{qerv}, and the FazyRV family, which provides scalable data-paths from 1- to 8-bit~\cite{2024_kissich_fazyrv}. For reference to a compact full‑width data-path design, PicoRV32~\cite{picorv32} is included as a 32‑bit baseline optimized for area. It serves as a comparison point for both the proposed 16‑bit serialized core and 32‑bit data-path implementation. For a fair area comparison, all state-of-the-art cores are synthesized without CSRs and with the same tool flow as the proposed SAILOR cores. The area results are shown in Table~\ref{tab:area_comparison_rv32i}.

Since the our architecture is designed to minimize area while still incorporating modules for improved performance and efficiency, the serialized RV32I SAILOR cores exhibit higher area requirements compared to their minimalist counterparts. For the 1‑bit data-path, our implementation results in an area increase of 7.74\,\% and 34.61\,\% relative to SERV and FazyRV, respectively. This overhead w.r.t. FazyRV decreases as the data-path width grows, reaching 19.76 \,\% at the 8‑bit configuration. In contrast, when compared to PicoRV32, our 16‑bit serialized and 32‑bit data-path cores achieve area reductions of 15.49\,\% and 21.26\,\%, respectively. These results indicate that the additional hardware required to improve performance and efficiency, such as an instruction fetch buffer and a dedicated load-/store-unit, does not incur substantial overhead compared to minimalist serialized designs. Furthermore, the observed area savings compared to conventional 32‑bit data-path designs highlight the suitability of our approach for resource‑constrained applications requiring a balanced area and performance trade-off.

These observations extend to cores with the \textit{Zkn–Zkt} cryptographic extensions enabled. The hardware overhead introduced by these extensions ranges from 5.54\,\% for the 16-bit serialized core relative to PicoRV32, up to 62.61\,\% for the 1-bit variant when compared to the corresponding 1-bit FazyRV implementation. The results reflect the inherent trade-offs between incorporating enhanced cryptographic functionality features and preserving compact area utilization across different data-path granularities.
Overall, the findings confirm that our architecture effectively bridges the gap between strictly area‑optimized serialized designs and conventional low‑area 32‑bit data-path cores, while enhancing cryptographic functionality at comparatively modest area costs.

\subsection{Area Comparison with State-of-the-Art Cryptographic RISC-V Cores}
\label{subsec:comparison_soa_crypto_riscv}

To underline the lightweightness of the proposed SAILOR cores, we compare their area with the state-of-the-art RISC-V cores implementing cryptography extensions similarly. The results are available in Table~\ref{tab:area_comparison_riscv_crypto}. Some state-of-the-art cores additionally implement instructions to support the ShangMi SM3 hash and SM4 block cipher \cite{riscv_scalar_crypto_v1.0.1}. Most of these implementations make use of a combined S-Box implementation for AES and SM4, targeting an efficient implementation for both ciphers. For a lightweight implementation, the proposed SAILOR cores employ a specialized area-constrained AES S-Box without encountering SM4 and only requiring 253.35~GE~\cite{2019_maximov_circuit_minimzation_smaller_faster_sboxes}.
The results show 35.59\,\% to 58.98\,\% smaller synthesis results compared to state-of-the-art. This demonstrates the overall lightweightness of the proposed cores, which can be deployed in resource-constrained IoT devices for cryptographic applications.

\input{tables/area_comparision_riscv_crypto}

%% file: diagrams/area_sailchip.tex

\pgfplotstableread{ 
Label rv32i rv32izbkbzkt rv32izbkczkt rv32izbkxzkt rv32izknezkt rv32izkndzkt rv32izknhzkt
1-Bit 12948.33322 279.6666742 366.0000113 201.3333321 532.6666667 3.6666642 1313.333366   
2-Bit 13090.99989 379.6666604 316.3333308 173.0000013 517.3333371 12 1250.000036   
4-Bit 13120.99989 277.3333459 295.6666654 294.0000063 514.6666692 6.0000012 1233.666703    
8-Bit 13179.66656 288.3333446 304.6666629 304.3333371 519.0000038 13.3333358 1267.33337
16-Bit 13278.66656 287.3333434 444.0000088 352.6666767 523.3333383 25.6666642 1275.333367
32-Bit 12647.99989 995.3333471 223 997.6666729 1160.666667 31 1446.666702

}\sailchipsyn

\begin{figure}[b]
\centering
\resizebox{\columnwidth}{!}{%
\begin{tikzpicture}

\begin{axis}[
    width=.09\textwidth, height=0.09\textwidth,
    ymin=0, ymax=1,     
    xtick=\empty,
    ytick={0},
    axis x line=bottom,
    axis y line=left,
    axis line style={-},    
    tick style={-}, 
]
\addplot[mark=none] coordinates{(0,0)};
\end{axis}

    \begin{axis}[
        ybar stacked,                       
        bar width=12pt,
        width=.5\textwidth,
        height=0.3\textwidth,
        symbolic x coords={1-Bit,2-Bit,4-Bit,8-Bit,16-Bit,32-Bit},
        xtick=data,
        ylabel={Area / GE},
        ymin=11000,
        axis y discontinuity=crunch,
        legend style={at={(0.5,-0.3)}, anchor=north, legend columns=3},
        ymajorgrids=true,
        xticklabel style={rotate=0},
        scaled y ticks=false,
        yticklabel style={
            /pgf/number format/.cd,
            fixed,
            fixed zerofill,
            precision=0,
            use comma
        }
    ]
    \addplot [fill=BrickRed] table [x=Label, y=rv32i] {\sailchipsyn};
    \addplot [fill=YellowGreen] table [x=Label, y=rv32izbkbzkt] {\sailchipsyn};
    \addplot [fill=Peach] table [x=Label, y=rv32izbkczkt] {\sailchipsyn};
    \addplot [fill=SpringGreen] table [x=Label, y=rv32izbkxzkt] {\sailchipsyn};
    \addplot [fill=RoyalBlue] table [x=Label, y=rv32izkndzkt] {\sailchipsyn};
    \addplot [fill=Goldenrod] table [x=Label, y=rv32izknezkt] {\sailchipsyn};
    \addplot [fill=SeaGreen] table [x=Label, y=rv32izknhzkt] {\sailchipsyn};
    \legend{
        RV32I,
        RV32I-Zbkb-Zkt,
        RV32I-Zbkc-Zkt,
        RV32I-Zbkx-Zkt,
        RV32I-Zknd-Zkt,
        RV32I-Zkne-Zkt,
        RV32I-Zknh-Zkt
    }
    \end{axis}

\end{tikzpicture}
}
\caption{Area Comparison of the SAILOR Serialized Cores}
\label{fig:area_comparison_endavour}
\end{figure}
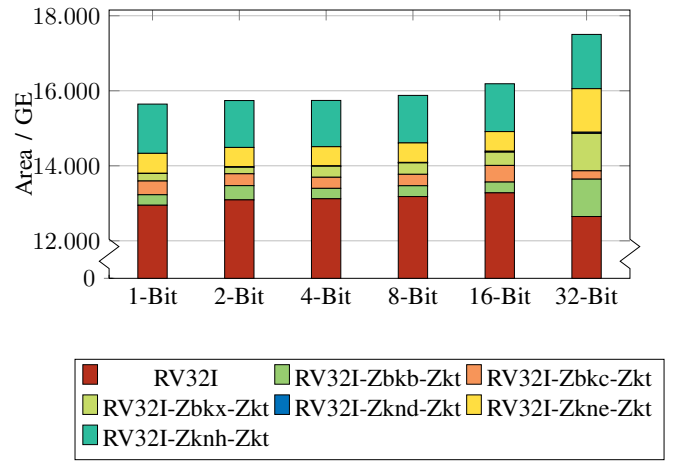

%% file: tables/area_comparison_rv32i.tex
\begin{table}[t]
\centering
\caption{Area Comparison with State-of-the-Art RV32I Cores}
\label{tab:area_comparison_rv32i}
\resizebox{\columnwidth}{!}{%
\begin{tabular}{|c|cccccc|}
\hline
\multirow{2}{*}{\textbf{Reference}} & \multicolumn{6}{c|}{\textbf{Area Consumption / kGE}}                                                                                                                                                           \\ \cline{2-7} 
                                    & \multicolumn{1}{c|}{\textbf{1-bit}} & \multicolumn{1}{c|}{\textbf{2-bit}} & \multicolumn{1}{c|}{\textbf{4-bit}} & \multicolumn{1}{c|}{\textbf{8-bit}} & \multicolumn{1}{c|}{\textbf{16-bit}} & \textbf{32-bit} \\ \hline
SERV~\cite{serv}                         & \multicolumn{1}{c|}{12.0}           & \multicolumn{1}{c|}{-}              & \multicolumn{1}{c|}{-}              & \multicolumn{1}{c|}{-}              & \multicolumn{1}{c|}{-}               & -               \\ \hline
QERV~\cite{qerv}                         & \multicolumn{1}{c|}{-}              & \multicolumn{1}{c|}{-}              & \multicolumn{1}{c|}{11.7}           & \multicolumn{1}{c|}{-}              & \multicolumn{1}{c|}{-}               & -               \\ \hline
FazyRV~\cite{2024_kissich_fazyrv}          & \multicolumn{1}{c|}{9.6}            & \multicolumn{1}{c|}{10.1}           & \multicolumn{1}{c|}{10.3}           & \multicolumn{1}{c|}{11.0}           & \multicolumn{1}{c|}{-}               & -               \\ \hline
PicoRV32~\cite{picorv32}                     & \multicolumn{1}{c|}{-}              & \multicolumn{1}{c|}{-}              & \multicolumn{1}{c|}{-}              & \multicolumn{1}{c|}{-}              & \multicolumn{1}{c|}{-}               & 15.3            \\ \hline
SAILOR RV32I                        & \multicolumn{1}{c|}{12.9}           & \multicolumn{1}{c|}{13.0}           & \multicolumn{1}{c|}{13.1}           & \multicolumn{1}{c|}{13.2}           & \multicolumn{1}{c|}{13.3}            & 12.6            \\ \hline
SAILOR RV32I-Zkn-Zkt                & \multicolumn{1}{c|}{15.6}           & \multicolumn{1}{c|}{15.7}           & \multicolumn{1}{c|}{15.7}           & \multicolumn{1}{c|}{15.9}           & \multicolumn{1}{c|}{16.2}            & 17.5            \\ \hline
\end{tabular}%
}
\end{table}

%% file: tables/area_comparision_riscv_crypto.tex
\begin{table}[t]
\centering
\caption{Area Comparison of RISC-V Architectures Implementing the Cryptographic Extension Suite}
\label{tab:area_comparison_riscv_crypto}
\resizebox{\columnwidth}{!}{%
\begin{tabular}{|c|cc|c|c|}
\hline
\multirow{2}{*}{Reference}                                       & \multicolumn{2}{c|}{Area / kGE}                & \multirow{2}{*}{Technology} & \multirow{2}{*}{Misc.}  \\ \cline{2-3}
                                                                 & \multicolumn{1}{c|}{Baseline}    & + Crypto    &                             &                           \\ \hline
\cite{2020_marshall_implementing_draft_crypto}                   & \multicolumn{1}{c|}{33.39}       & 39.25       & N/A                         & Draft Extension, +SM3/SM4 \\ \hline
\cite{2023_gewehr_improving_efficiency_cryptographic_algorithms} & \multicolumn{1}{c|}{34,43}       & 38.13       & 28nm Commercial             & Only Zkne \& Zknh         \\ \hline
\cite{2022_nisanci_symmetric_crypto_riscv}                       & \multicolumn{1}{c|}{19.71}       & 27.17       & N/A                         & +SM4                      \\ \hline
{[}This work{]}                                                  & \multicolumn{1}{c|}{$\leq$13.28} & $\leq$17.50 & 45nm Open-Cell              & Full Zkn support          \\ \hline
\end{tabular}%
}
\end{table}

%% file: sections/05_results.tex
\section{Performance-, Energy- and Effciency-Evaluation}
\label{sec:evaluation}

\input{tables/code_size_comparison_crypto_algorithms}

This section presents the evaluation of the performance, energy, and efficiency of the proposed SAILOR cores and compares them to state-of-the-art serialized and compact 32-bit data-path cores. To perform the analysis, we employ the cryptographic benchmark suite from~\cite{riscv_crypto_benchmarks}. The benchmark set comprises AES-128/-192/-256 block cipher encryption and decryption~\cite{aes_nist_standard}, SHA-256/-512 hash algorithms~\cite{sha2_nist_standard}, and the Prince block cipher~\cite{prince_block_cipher} S-Box computation performed through permutation operations. The benchmarks comprise an RV32I baseline and \textit{Zkn} extended variant, which are compiled using riscv-unknown-elf-gcc~14.2.0~\cite{riscv_gcc} with \textit{{-O3}} optimization flag. Cycle-accurate simulation and VCD generation are performed utilizing the \textit{chiseltest}~\cite{chiseltest} simulation framework and Synopsys VCS 2022.06-SP2~\cite{synopsys_vcs}, while detailed, benchmark-specific energy analysis is obtained through Synopsys Prime Power 2024.09-SP2~\cite{synopsys_primepower}.

Table~\ref{tab:code_size_comparison_crypto} shows the code size comparison between the RV32I and Zkn extended code. The results show code size reduction from the RV32I to the Zkn compiled code of 53.13\,\% to 78.51\,\%. The AES algorithm software benefits the most from the cryptographic extension. In particular, the instructions required to implement the S-Box T-table implementation are embedded in the middle- and end-round instructions, contributing to reducing the instruction count and likewise code size. For architectural comparisons against state-of-the-art cores, we additionally employ Coremark~\cite{2009_coremark} and Dhrystone~\cite{2002_dhrystone}, both compiled for the RV32I ISA under the same compiler optimizations. 

In the following, our evaluation focuses on speedup, energy reduction, and improvements in efficiency metrics such as area–time and energy–delay products, thereby quantifying performance and efficiency gains across the considered benchmark workloads. We compare the proposed cores to the state-of-the-art serialized and compact 32-bit
path cores and analyze relative improvements w.r.t. the metrics. For detailed benchmark results please refer to Table~\ref{tab:aes_128_encryption_benchmark}--\ref{tab:dhrystone_benchmark} in the Appendix.

\subsection{Speedup Comparison}
\label{subsubsec:speedup_comparison}

Fig.~\ref{fig:speedup_comparison} presents the speedup comparison of all evaluated cores, using PicoRV32~\cite{picorv32} as the reference design. For the 1‑, 2‑, 4‑, and 8‑bit serialized designs, the proposed SAILOR cores with the RV32I ISA achieve a 1.7–2.3$\times$ speedup across benchmarks compared to competitor architectures with the same serialization width. When comparing the proposed 16‑bit serialized and 32‑bit data-path cores against the state‑of‑the‑art 32‑bit data-path core, a performance improvement of 2.0–3.8$\times$ is observed. Notably, the 16‑bit serialized SAILOR core outperforms the 32‑bit data-path implementation of SAILOR by 12.5–60.9\,\% across benchmarks. Strong gains occur in Coremark due to the efficient utilization of the 16‑bit serialized shift unit. The performance crossover point relative to PicoRV32 occurs at the proposed 4‑bit serialized core followed by the 8‑bit variant, which delivers up to 2.3$\times$ higher performance. These findings underline the targeted performance enhancements within area‑optimized serial data-path architectures, while retaining compact area as analyzed in Section~\ref{subsec:area_comparison_soa_rv32i}.

\input{diagrams/speedup_comparision}

In the case of the proposed \textit{Zkn–Zkt} extended SAILOR cores, the serialized designs achieve a 6.0–11.0$\times$ speedup compared to their respective serialized state-of-the-art counterparts. Likewise, the proposed 16‑bit serialized and 32‑bit data-path SAILOR cores with the \textit{Zkn–Zkt} extensions deliver a 5.2–12.0$\times$ performance improvement relative to PicoRV32. Overall, all \textit{Zkn–Zkt} enhanced cores show performance gains for cryptographic workloads compared to PicoRV32. However, for benchmarks that do not require hardware support for cryptographic algorithms, such as Coremark and Dhrystone, the constant‑time execution semantics impose a performance reduction of up to 55\,\%. This observation highlights an optimization opportunity for non‑security‑critical applications, where instructions operating on sensitive data would maintain constant‑time behavior, while computations on non‑security-critical data could be optimized for higher performance. Such flexibility can be realized through run‑time configuration, allowing the core to adapt its functionality to the specific security and performance requirements of a given application. 

The results demonstrate that targeted adaptations of typically highly area‑optimized core architectures can substantially improve the performance of cryptographic applications while maintaining a compact area footprint.

\subsection{Energy Consumption Comparison}
\label{subsubsec:energy_reduction_comparison}

Fig.~\ref{fig:energy_reduction_comparison} illustrates the energy reduction across all benchmarks, using PicoRV32 as the reference design. Consistent with the speedup improvements analyzed in Section~\ref{subsubsec:speedup_comparison}, the proposed 1‑, 2‑, 4‑, and 8‑bit serialized SAILOR cores without the \textit{Zkn–Zkt} extension achieve a 1.92–2.82$\times$ reduction in energy compared to state-of-the-art competitor designs with the same serialization width. Similarly, the proposed 16‑bit serialized and 32‑bit data-path SAILOR cores show a 2.26–3.69$\times$ energy reduction relative to PicoRV32. Except for the SHA‑256 and SHA‑512 hash benchmarks, the 16‑bit serialized core further outperforms the proposed 32‑bit data-path core by 7.40–35.16\,\% in energy reduction. The deviation observed in SHA‑256 and SHA‑512 stems from their shift‑operation dominance: while shifts execute on average faster in the proposed 16‑bit architecture, their execution results in higher power and, therefore, energy consumption. Across most benchmarks, energy reductions w.r.t. PicoRV32 are achieved from the 2‑bit serialized SAILOR core, whereas the 4‑bit serialized variant achieves up to 1.8$\times$ energy savings. These results highlight how architectural trade‑offs in small‑area designs can simultaneously reduce energy consumption while maintaining a compact area footprint.

\input{diagrams/energy_reduction_comparison}

By integrating dedicated hardware support for the RISC‑V cryptography extensions, energy consumption can be further reduced. The enhanced SAILOR cores achieve a 2.75–4.62$\times$ energy reduction compared to their corresponding proposed RV32I baseline cores. Overall, the proposed SAILOR cores with hardware support for the Zkn-Zkt extensions provide energy savings of up to 12.79$\times$ relative to PicoRV32. A noteworthy observation is that SHA‑256 and SHA‑512 strongly benefit from wider serialization widths. For example, the 32‑bit \textit{Zkn–Zkt} core achieves a 13.24$\times$ energy reduction compared to its 1‑bit counterpart. The primary factor contributing to this improvement is the reduction in clock cycles required for logical operations within the ALU, while shift operations for these workloads execute with single‑cycle latency. For applications where such workloads dominate, this points to further optimization potential: in addition to fixed shifts and rotations, implementing dedicated XOR operations could reduce execution cycles and energy consumption further, albeit at the expense of additional area.

\subsection{Efficiency Comparison}
\label{subsubsec:efficiency_analysis}

To quantify the efficiency improvements of the proposed modular and scalable architecture, we analyze the energy–delay product (EDP) and area–time product (ATP) metrics. 

The EDP for a given benchmark $Bench_i$ executed on a $CPU_j$ can be calculated as:

\begin{equation}
    EDP_{Bench_{i},CPU_{j}} = E_{Bench_{i},CPU_{j}} \times T_{Bench_{i},CPU_{j}}, \text{where}
\end{equation}

$E_{Bench_{i},CPU_{j}}$ is the energy consumed for the benchmark executed on $CPU_j$ and $T_{Bench_{i},CPU_{j}}$ is the execution time, respectively. A lower EDP indicates efficiency improvements when comparing the same benchmark executed on different cores. Likewise, the ATP for a given benchmark $Bench_i$ executed on a $CPU_j$ is derived as:

\begin{equation}
    ATP_{Bench_{i},CPU_{j}} = A_{CPU_{j}} \times T_{Bench_{i},CPU_{j}}, \text{where}
\end{equation}

$A_{CPU_{j}}$ is the area footprint of $CPU_j$. Similarly, a lower ATP indicates efficiency improvements when comparing the same benchmark executed on different cores.

\input{diagrams/edp_comparison}

The EDP improvement results are shown in Fig.~\ref{fig:edp_reduction_comparison}.
The performance analysis in Section~\ref{subsubsec:speedup_comparison} and the energy analysis in Section~\ref{subsubsec:energy_reduction_comparison} both demonstrate consistent improvements across the considered SAILOR core configurations. Consequently, the EDP for the proposed serialized SAILOR cores shows a 4.0–4.7$\times$ improvement over state-of-the-art serialized cores, and up to an 11$\times$ enhancement for the 16‑ and 32‑bit versions relative to PicoRV32. Notably, implementations with enabled cryptography extensions exhibit EDP improvements of up to 153$\times$. These results underline the energy efficiency of the proposed architecture, especially in the context of cryptographic applications.

\input{diagrams/atp_comparison}

Although the proposed architectures require more area compared to the state-of-the-art, as analyzed in Section~\ref{subsec:area_comparison_soa_rv32i}, the ATP analysis shows significant improvement.
The results depicted in Fig.~\ref{fig:atp_reduction_comparison} show 1.50-1.96$\times$ enhancement for the proposed serialized SAILOR cores with the RV32I baseline ISA and 3.9-6.8$\times$ improvement with configured \textit{Zkn-Zkt} cryptography extension compared to their state-of-the-art serialized counterparts.
Likewise, the proposed 16-bit serialized and 32-bit data-path version improves the ATP by 2.40-3.05$\times$ for the baseline RV32I ISA and 3.59-10.51$\times$ with enabled cryptography extensions compared to PicoRV32.

As also observed in Section~\ref{subsubsec:speedup_comparison} for performance, the ATP improvement is reduced by 18.75\,\% to 27.7\,\% for the Coremark and Dhrystone benchmarks when the cryptography extensions are enabled. This reduction is primarily due to the data-independent constant-time execution enforced by the \textit{Zkt} extension. This observation suggests optimization potential by selectively disabling constant-time execution when operating on non-security-critical data. In total, the ATP improvement demonstrates the architectural area efficiency of our design not only in general terms but particularly for cryptographic applications. 

%% file: tables/code_size_comparison_crypto_algorithms.tex
\begin{table}[t]
\centering
\caption{Code Size Comparison of Cryptographic Algorithms with RV32I ISA and enabled -Zkn Extension }
\label{tab:code_size_comparison_crypto}
\resizebox{\columnwidth}{!}{%
\begin{tabular}{|c|ccc|ccc|}
\hline
\multirow{2}{*}{Benchmark} & \multicolumn{3}{c|}{RV32I / bytes}                             & \multicolumn{3}{c|}{+Zkn / bytes}                              \\ \cline{2-7} 
                           & \multicolumn{1}{c|}{.data} & \multicolumn{1}{c|}{.bss} & .text & \multicolumn{1}{c|}{.data} & \multicolumn{1}{c|}{.bss} & .text \\ \hline
AES128 Enc.                & \multicolumn{1}{c|}{80}    & \multicolumn{1}{c|}{432}  & 4764  & \multicolumn{1}{c|}{90}    & \multicolumn{1}{c|}{432}  & 612   \\ \hline
AES128 Dec.                & \multicolumn{1}{c|}{80}    & \multicolumn{1}{c|}{432}  & 4620  & \multicolumn{1}{c|}{90}    & \multicolumn{1}{c|}{432}  & 704   \\ \hline
AES192 Enc.                & \multicolumn{1}{c|}{80}    & \multicolumn{1}{c|}{496}  & 4764  & \multicolumn{1}{c|}{90}    & \multicolumn{1}{c|}{496}  & 636   \\ \hline
AES192 Dec.                & \multicolumn{1}{c|}{80}    & \multicolumn{1}{c|}{496}  & 4620  & \multicolumn{1}{c|}{90}    & \multicolumn{1}{c|}{496}  & 728   \\ \hline
AES256 Enc.                & \multicolumn{1}{c|}{80}    & \multicolumn{1}{c|}{560}  & 4764  & \multicolumn{1}{c|}{90}    & \multicolumn{1}{c|}{560}  & 688   \\ \hline
AES256 Dec.                & \multicolumn{1}{c|}{80}    & \multicolumn{1}{c|}{560}  & 4620  & \multicolumn{1}{c|}{90}    & \multicolumn{1}{c|}{560}  & 780   \\ \hline
SHA-256                    & \multicolumn{1}{c|}{320}   & \multicolumn{1}{c|}{32}   & 6476  & \multicolumn{1}{c|}{320}   & \multicolumn{1}{c|}{32}   & 2848  \\ \hline
SHA-512                    & \multicolumn{1}{c|}{1216}  & \multicolumn{1}{c|}{64}   & 15024 & \multicolumn{1}{c|}{1216}  & \multicolumn{1}{c|}{64}   & 7156  \\ \hline
Permutation                & \multicolumn{1}{c|}{16}    & \multicolumn{1}{c|}{-}    & 2488  & \multicolumn{1}{c|}{16}    & \multicolumn{1}{c|}{-}    & 540   \\ \hline
\end{tabular}%
}
\end{table}

%% file: diagrams/speedup_comparision.tex
\begin{figure}[t]
\centering
\resizebox{\columnwidth}{!}{%
\begin{tikzpicture}

\pgfplotstableread{diagrams/data/cryptobenchmark_speedup_3d_grid.dat}\tabledata

\begin{axis}[
    width=1.3\columnwidth, 
    view={-50}{80},
    view/h=-30,
    plot box ratio=1 1.5 5,
    colorbar,
    point meta min=0,
    point meta max=12,
    xlabel={Processor (CPU)},
    ylabel={Benchmark},
    zlabel={Speedup},
    xtick={1,...,19},
    xticklabels={
        PicoRV32, 
        SERV, 
        QERV, 
        FazyRV 1-bit, 
        FazyRV 2-bit, 
        FazyRV 4-bit, 
        FazyRV 8-bit, 
        SAILOR 1-bit RV32I,
        SAILOR 2-bit RV32I,
        SAILOR 4-bit RV32I,
        SAILOR 8-bit RV32I,
        SAILOR 16-bit RV32I,
        SAILOR 32-bit RV32I,
        SAILOR 1-bit Zkn-Zkt,
        SAILOR 2-bit Zkn-Zkt,
        SAILOR 4-bit Zkn-Zkt,
        SAILOR 8-bit Zkn-Zkt,
        SAILOR 16-bit Zkn-Zkt,
        SAILOR 32-bit Zkn-Zkt
    },
    x tick label style={font=\small, rotate=-45, anchor=west},
    xtick distance=1,
    ytick={1,...,11},
    yticklabels={
        AES-128-Encryption,
        AES-128-Decryption,
        AES-192-Encryption,
        AES-192-Decryption,
        AES-256-Encryption,
        AES-256-Decryption,
        SHA-256,
        SHA-512,
        Permutation,
        Coremark,
        Dhrystone
    },
    y tick label style={font=\small, anchor=east, rotate=45},
    y dir=reverse,
    ztick distance=2,
    grid=major,
]
\addplot3 [
    surf,
    scatter,
    fill=white,
    opacity=0.7,
    mesh/ordering=y varies,
    mesh/rows=11,
] table {\tabledata};
\end{axis}
\end{tikzpicture}
}
\caption{CPU-Speedup Comparison}
\label{fig:speedup_comparison}
\end{figure}

%% file: diagrams/energy_reduction_comparison.tex
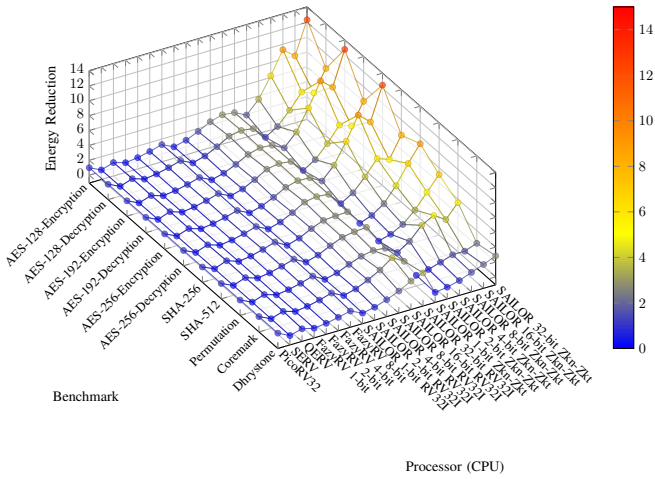
\begin{figure}[t]
\resizebox{\columnwidth}{!}{%
\begin{tikzpicture}

\pgfplotstableread{diagrams/data/cryptobenchmark_energy_reduction_3d_grid.dat}\tabledata

\begin{axis}[
    width=1.3\columnwidth, 
    view={-50}{80},
    view/h=-30,
    plot box ratio=1 1.5 5,
    colorbar,
    point meta min=0,
    point meta max=15,
    xlabel={Processor (CPU)},
    ylabel={Benchmark},
    zlabel={Energy Reduction},
    xtick={1,...,19},
    xticklabels={
        PicoRV32, 
        SERV, 
        QERV, 
        FazyRV 1-bit, 
        FazyRV 2-bit, 
        FazyRV 4-bit, 
        FazyRV 8-bit, 
        SAILOR 1-bit RV32I,
        SAILOR 2-bit RV32I,
        SAILOR 4-bit RV32I,
        SAILOR 8-bit RV32I,
        SAILOR 16-bit RV32I,
        SAILOR 32-bit RV32I,
        SAILOR 1-bit Zkn-Zkt,
        SAILOR 2-bit Zkn-Zkt,
        SAILOR 4-bit Zkn-Zkt,
        SAILOR 8-bit Zkn-Zkt,
        SAILOR 16-bit Zkn-Zkt,
        SAILOR 32-bit Zkn-Zkt
    },
    x tick label style={font=\small, rotate=-45, anchor=west},
    xtick distance=1,
    ytick={1,...,11},
    yticklabels={
        AES-128-Encryption,
        AES-128-Decryption,
        AES-192-Encryption,
        AES-192-Decryption,
        AES-256-Encryption,
        AES-256-Decryption,
        SHA-256,
        SHA-512,
        Permutation,
        Coremark,
        Dhrystone
    },
    y tick label style={font=\small, anchor=east, rotate=45},
    y dir=reverse,
    ztick distance=2,
    grid=major,
]
\addplot3 [
    surf,
    scatter,
    fill=white,
    opacity=0.7,
    mesh/ordering=y varies,
    mesh/rows=11,
] table {\tabledata};
\end{axis}
\end{tikzpicture}
}
\caption{Energy-Reduction Comparison}
\label{fig:energy_reduction_comparison}
\end{figure}

%% file: diagrams/edp_comparison.tex
\begin{figure}[t]
\resizebox{\columnwidth}{!}{%
\begin{tikzpicture}

\pgfplotstableread{diagrams/data/cryptobenchmark_edp_3d_grid.dat}\tabledata

\begin{axis}[
    width=1.3\columnwidth, 
    view={-50}{80},
    view/h=-30,
    plot box ratio=1 1.5 5,
    colorbar,
    point meta min=0,
    point meta max=144,
    xlabel={Processor (CPU)},
    ylabel={Benchmark},
    zlabel={EDP Improvement},
    xtick={1,...,19},
    xticklabels={
        PicoRV32, 
        SERV, 
        QERV, 
        FazyRV 1-bit, 
        FazyRV 2-bit, 
        FazyRV 4-bit, 
        FazyRV 8-bit, 
        SAILOR 1-bit RV32I,
        SAILOR 2-bit RV32I,
        SAILOR 4-bit RV32I,
        SAILOR 8-bit RV32I,
        SAILOR 16-bit RV32I,
        SAILOR 32-bit RV32I,
        SAILOR 1-bit Zkn-Zkt,
        SAILOR 2-bit Zkn-Zkt,
        SAILOR 4-bit Zkn-Zkt,
        SAILOR 8-bit Zkn-Zkt,
        SAILOR 16-bit Zkn-Zkt,
        SAILOR 32-bit Zkn-Zkt
    },
    x tick label style={font=\small, rotate=-45, anchor=west},
    xtick distance=1,
    ytick={1,...,11},
    yticklabels={
        AES-128-Encryption,
        AES-128-Decryption,
        AES-192-Encryption,
        AES-192-Decryption,
        AES-256-Encryption,
        AES-256-Decryption,
        SHA-256,
        SHA-512,
        Permutation,
        Coremark,
        Dhrystone
    },
    y tick label style={font=\small, anchor=east, rotate=45},
    y dir=reverse,
    ztick distance=20,
    grid=major,
]
\addplot3 [
    surf,
    scatter,
    fill=white,
    opacity=0.7,
    mesh/ordering=y varies,
    mesh/rows=11,
] table {\tabledata};
\end{axis}
\end{tikzpicture}
}
\caption{Energy-Delay Product Improvement Comparison}
\label{fig:edp_reduction_comparison}
\end{figure}
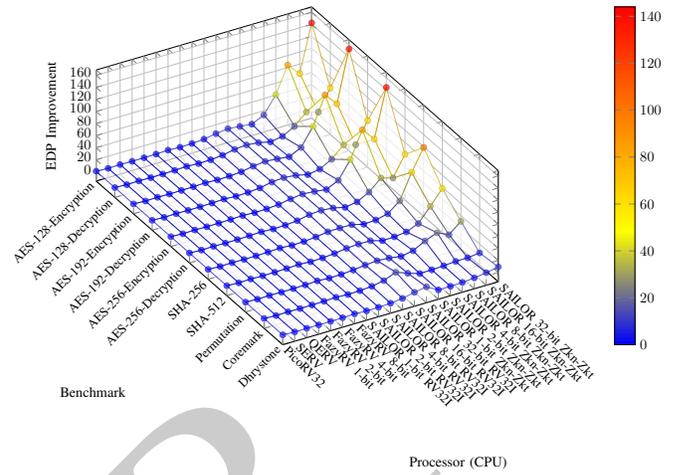

%% file: diagrams/atp_comparison.tex
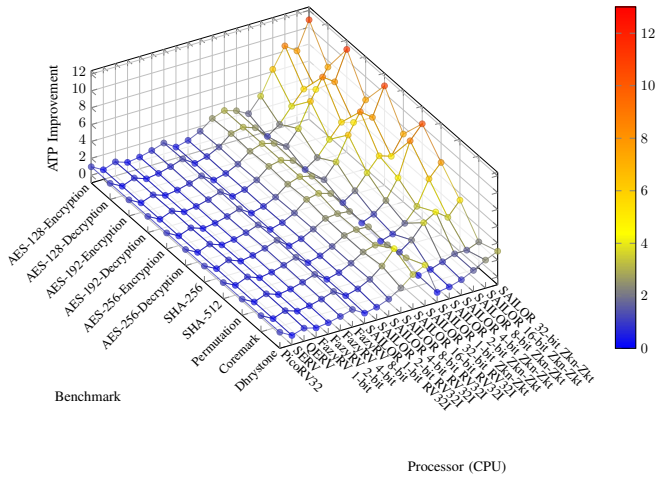
\begin{figure}[t]
\resizebox{\columnwidth}{!}{%
\begin{tikzpicture}

\pgfplotstableread{diagrams/data/cryptobenchmark_atp_3d_grid.dat}\tabledata

\begin{axis}[
    width=1.3\columnwidth, 
    view={-50}{80},
    view/h=-30,
    plot box ratio=1 1.5 5,
    colorbar,
    point meta min=0,
    point meta max=13,
    xlabel={Processor (CPU)},
    ylabel={Benchmark},
    zlabel={ATP Improvement},
    xtick={1,...,19},
    xticklabels={
        PicoRV32, 
        SERV, 
        QERV, 
        FazyRV 1-bit, 
        FazyRV 2-bit, 
        FazyRV 4-bit, 
        FazyRV 8-bit, 
        SAILOR 1-bit RV32I,
        SAILOR 2-bit RV32I,
        SAILOR 4-bit RV32I,
        SAILOR 8-bit RV32I,
        SAILOR 16-bit RV32I,
        SAILOR 32-bit RV32I,
        SAILOR 1-bit Zkn-Zkt,
        SAILOR 2-bit Zkn-Zkt,
        SAILOR 4-bit Zkn-Zkt,
        SAILOR 8-bit Zkn-Zkt,
        SAILOR 16-bit Zkn-Zkt,
        SAILOR 32-bit Zkn-Zkt
    },
    x tick label style={font=\small, rotate=-45, anchor=west},
    xtick distance=1,
    ytick={1,...,11},
    yticklabels={
        AES-128-Encryption,
        AES-128-Decryption,
        AES-192-Encryption,
        AES-192-Decryption,
        AES-256-Encryption,
        AES-256-Decryption,
        SHA-256,
        SHA-512,
        Permutation,
        Coremark,
        Dhrystone
    },
    y tick label style={font=\small, anchor=east, rotate=45},
    y dir=reverse,
    ztick distance=2,
    grid=major,
]
\addplot3 [
    surf,
    scatter,
    fill=white,
    opacity=0.7,
    mesh/ordering=y varies,
    mesh/rows=11,
] table {\tabledata};
\end{axis}
\end{tikzpicture}
}
\caption{Area-Time Product Improvement Comparison}
\label{fig:atp_reduction_comparison}
\end{figure}

%% file: sections/06_conclusion.tex
\section{Conclusion}
\label{sec:conclusion}

In this paper, we presented SAILOR, a scalable and energy-efficient ultra-lightweight RISC-V architecture for IoT security applications. Its configurable execution data-path scales from 1 to 32 bits, balancing performance and energy requirements for embedded IoT devices. By reusing existing functionality, SAILOR integrates cryptographic extensions with minimal hardware overhead while supporting NIST-standard cryptography (Zkn-Zkt), achieving significant energy savings and performance gains.
Implemented as a proof of concept, SAILOR is comprehensively evaluated for area, energy, and efficiency trade-offs. It fills a critical gap in RISC-V research by targeting resource-constrained IoT devices and demonstrates that lightweight cryptographic extensions can be integrated without prohibitive overhead, and that energy- or area-optimized designs need not compromise performance. Its results surpass state-of-the-art solutions, confirming the architecture’s practicality for real-world IoT deployments.
\vfill\eject

%% file: sections/07_appendix.tex
\appendix[Detailed Benchmark Results] 

\input{tables/aes_128_encryption_benchmark_results}

\input{tables/aes_128_decryption_benchmark_results}

\input{tables/aes_192_encryption_benchmark_results}

\input{tables/aes192_decryption_benchmark_results}

\input{tables/aes_256_encryption_benchmark_results}

\input{tables/aes_256_decryption_benchmark_results}

\input{tables/sha_256_benchmark_results}

\input{tables/sha_512_benchmark_results}

\input{tables/permutation_benchmark_results}

\input{tables/coremark_benchmark_results}

\input{tables/dhrystone_benchmark_results}

%% file: tables/aes_128_encryption_benchmark_results.tex
\begin{table}[H]
\centering
\caption{Comparison of AES-128 Encryption}
\label{tab:aes_128_encryption_benchmark}
\resizebox{\columnwidth}{!}{%
\begin{tabular}{|cc|c|c|c|c|}
\hline
\multicolumn{2}{|c|}{\textbf{Processor}}                                                                                       & \textbf{Execution Time / $\mu$s} & \textbf{$E_{average}$ / nJ} & \textbf{Energy Delay Product} & \textbf{Area Time Product} \\ \hline
\multicolumn{2}{|c|}{\textbf{PicoRV32}}                                                                                        & 704,89                           & 563,35                      & 397098                        & 10,81                      \\ \hline
\multicolumn{2}{|c|}{\textbf{SERV}}                                                                                            & 4063,93                          & 2489,56                     & 10117412                      & 48,84                      \\ \hline
\multicolumn{2}{|c|}{\textbf{QERV}}                                                                                            & 1243,31                          & 758,92                      & 943568                        & 14,59                      \\ \hline
\multicolumn{1}{|c|}{\multirow{4}{*}{\textbf{Fazy}}}                                                                  & 1-bit  & 4281,20                          & 2271,18                     & 9723361                       & 41,18                      \\ \cline{2-6} 
\multicolumn{1}{|c|}{}                                                                                                & 2-bit  & 2267,27                          & 1250,63                     & 2835507                       & 22,83                      \\ \cline{2-6} 
\multicolumn{1}{|c|}{}                                                                                                & 4-bit  & 1260,10                          & 700,87                      & 883163                        & 13,00                      \\ \cline{2-6} 
\multicolumn{1}{|c|}{}                                                                                                & 8-bit  & 756,50                           & 448,45                      & 339255                        & 8,33                       \\ \hline
\multicolumn{1}{|c|}{\multirow{6}{*}{\textbf{\begin{tabular}[c]{@{}c@{}}SAILOR\\ RV32I\,/\\ -Zkn-Zkt\end{tabular}}}} & 1-bit  & 2067,30\,/\,373,10               & 1176,50\,/\,296,32          & 2432179\,/\,110556            & 26,77\,/\,5,88             \\ \cline{2-6} 
\multicolumn{1}{|c|}{}                                                                                                & 2-bit  & 1070,82\,/\,191,92               & 634,35\,/\,159,43           & 679279\,/\,30597              & 14,02\,/\,3,09             \\ \cline{2-6} 
\multicolumn{1}{|c|}{}                                                                                                & 4-bit  & 572,58\,/\,107,48                & 354,60\,/\,90,56            & 203036\,/\,9734               & 7,51\,/\,1,73              \\ \cline{2-6} 
\multicolumn{1}{|c|}{}                                                                                                & 8-bit  & 323,46\,/\,77,56                 & 231,18\,/\,60,92            & 74776\,/\,4725                & 4,26\,/\,1,25              \\ \cline{2-6} 
\multicolumn{1}{|c|}{}                                                                                                & 16-bit & 266,56\,/\,87,20                 & 195,07\,/\,70,82            & 51997\,/\,6176                & 3,54\,/\,1,42              \\ \cline{2-6} 
\multicolumn{1}{|c|}{}                                                                                                & 32-bit & 327,88\,/\,61,01                 & 209,06\,/\,45,70            & 68545\,/\,2788                & 4,15\,/\,0,99              \\ \hline
\end{tabular}%
}
\end{table}

%% file: tables/aes_128_decryption_benchmark_results.tex
\begin{table}[H]
\centering
\caption{Comparison of AES-128 Decryption}
\label{tab:aes_128_decryption_benchmark}
\resizebox{\columnwidth}{!}{%
\begin{tabular}{|cc|c|c|c|c|}
\hline
\multicolumn{2}{|c|}{\textbf{Processor}}                                                                                       & \textbf{Execution Time / $\mu$s} & \textbf{$E_{average}$ / nJ} & \textbf{Energy Delay Product} & \textbf{Area Time Product} \\ \hline
\multicolumn{2}{|c|}{\textbf{PicoRV32}}                                                                                        & 1139,16                          & 910,76                      & 1037500                       & 17,47                      \\ \hline
\multicolumn{2}{|c|}{\textbf{SERV}}                                                                                            & 6179,73                          & 3772,73                     & 23314423                      & 74,27                      \\ \hline
\multicolumn{2}{|c|}{\textbf{QERV}}                                                                                            & 1886,71                          & 1148,82                     & 2167486                       & 22,14                      \\ \hline
\multicolumn{1}{|c|}{\multirow{4}{*}{\textbf{Fazy}}}                                                                  & 1-bit  & 6562,84                          & 3461,24                     & 22715576                      & 63,13                      \\ \cline{2-6} 
\multicolumn{1}{|c|}{}                                                                                                & 2-bit  & 3474,83                          & 1908,03                     & 6630077                       & 34,99                      \\ \cline{2-6} 
\multicolumn{1}{|c|}{}                                                                                                & 4-bit  & 1928,50                          & 1068,39                     & 2060388                       & 19,90                      \\ \cline{2-6} 
\multicolumn{1}{|c|}{}                                                                                                & 8-bit  & 1155,12                          & 684,99                      & 791241                        & 12,71                      \\ \hline
\multicolumn{1}{|c|}{\multirow{6}{*}{\textbf{\begin{tabular}[c]{@{}c@{}}SAILOR\\ RV32I\,/\\ -Zkn-Zkt\end{tabular}}}} & 1-bit  & 3150,50\,/\,999,58               & 1702,85\,/\,806,56          & 5364814\,/\,806222            & 40,79\,/\,15,76            \\ \cline{2-6} 
\multicolumn{1}{|c|}{}                                                                                                & 2-bit  & 1629,86\,/\,511,84               & 919,40\,/\,432,81           & 1498500\,/\,221530            & 21,34\,/\,8,23             \\ \cline{2-6} 
\multicolumn{1}{|c|}{}                                                                                                & 4-bit  & 869,54\,/\,282,76                & 514,59\,/\,239,36           & 447460\,/\,67680              & 11,41\,/\,4,54             \\ \cline{2-6} 
\multicolumn{1}{|c|}{}                                                                                                & 8-bit  & 489,38\,/\,197,80                & 337,62\,/\,157,43           & 165226\,/\,31139              & 6,45\,/\,3,20              \\ \cline{2-6} 
\multicolumn{1}{|c|}{}                                                                                                & 16-bit & 432,16\,/\,214,48                & 303,59\,/\,175,32           & 131200\,/\,37602              & 5,74\,/\,3,50              \\ \cline{2-6} 
\multicolumn{1}{|c|}{}                                                                                                & 32-bit & 540,98\,/\,148,31                & 329,24\,/\,113,78           & 178112\,/\,16875              & 6,84\,/\,2,42              \\ \hline
\end{tabular}%
}
\end{table}

%% file: tables/aes_192_encryption_benchmark_results.tex
\begin{table}[H]
\centering
\caption{Comparison of AES-192 Encryption}
\label{tab:aes_192_encryption_benchmark}
\resizebox{\columnwidth}{!}{%
\begin{tabular}{|cc|c|c|c|c|}
\hline
\multicolumn{2}{|c|}{\textbf{Processor}}                                                                                       & \textbf{Execution Time / $\mu$s} & \textbf{$E_{average}$ / nJ} & \textbf{Energy Delay Product} & \textbf{Area Time Product} \\ \hline
\multicolumn{2}{|c|}{\textbf{PicoRV32}}                                                                                        & 814,10                           & 650,79                      & 529809                        & 12,49                      \\ \hline
\multicolumn{2}{|c|}{\textbf{SERV}}                                                                                            & 4598,65                          & 2819,43                     & 12965582                      & 55,27                      \\ \hline
\multicolumn{2}{|c|}{\textbf{QERV}}                                                                                            & 1408,59                          & 859,52                      & 1210714                       & 16,53                      \\ \hline
\multicolumn{1}{|c|}{\multirow{4}{*}{\textbf{Fazy}}}                                                                  & 1-bit  & 4840,59                          & 2569,39                     & 12437340                      & 46,56                      \\ \cline{2-6} 
\multicolumn{1}{|c|}{}                                                                                                & 2-bit  & 2564,58                          & 1412,83                     & 3623308                       & 25,82                      \\ \cline{2-6} 
\multicolumn{1}{|c|}{}                                                                                                & 4-bit  & 1426,25                          & 794,99                      & 1133857                       & 14,72                      \\ \cline{2-6} 
\multicolumn{1}{|c|}{}                                                                                                & 8-bit  & 857,07                           & 508,50                      & 435820                        & 9,43                       \\ \hline
\multicolumn{1}{|c|}{\multirow{6}{*}{\textbf{\begin{tabular}[c]{@{}c@{}}SAILOR\\ RV32I\,/\\ -Zkn-Zkt\end{tabular}}}} & 1-bit  & 2352,14\,/\,422,37               & 1338,37\,/\,337,01          & 3148028\,/\,142343            & 30,46\,/\,6,66             \\ \cline{2-6} 
\multicolumn{1}{|c|}{}                                                                                                & 2-bit  & 1218,22\,/\,217,17               & 721,92\,/\,181,68           & 879454\,\,39456               & 15,95\,/\,3,49             \\ \cline{2-6} 
\multicolumn{1}{|c|}{}                                                                                                & 4-bit  & 651,26\,/\,121,41                & 403,26\,/\,102,66           & 262627\,/\,12464              & 8,55\,/\,1,95              \\ \cline{2-6} 
\multicolumn{1}{|c|}{}                                                                                                & 8-bit  & 367,78\,/\,87,21                 & 262,89\,/\,68,48            & 96685\,/\,5972                & 4,85\,/\,1,41              \\ \cline{2-6} 
\multicolumn{1}{|c|}{}                                                                                                & 16-bit & 307,44\,/\,97,47                 & 224,40\,/\,79,37            & 68990\,/\,7736                & 4,08\,/\,1,59              \\ \cline{2-6} 
\multicolumn{1}{|c|}{}                                                                                                & 32-bit & 369,54\,/\,67,87                 & 235,36\,/\,50,87            & 86975\,/\,3452                & 4,67\,/\,1,11              \\ \hline
\end{tabular}%
}
\end{table}

%% file: tables/aes192_decryption_benchmark_results.tex
\begin{table}[H]
\centering
\caption{Comparison of AES-192 Decryption}
\label{tab:aes_192_decryption_benchmark}
\resizebox{\columnwidth}{!}{%
\begin{tabular}{|cc|c|c|c|c|}
\hline
\multicolumn{2}{|c|}{\textbf{Processor}}                                                                                       & \textbf{Execution Time / $\mu$s} & \textbf{$E_{average}$ / nJ} & \textbf{Energy Delay Product} & \textbf{Area Time Product} \\ \hline
\multicolumn{2}{|c|}{\textbf{PicoRV32}}                                                                                        & 1346,19                          & 1075,34                     & 1447607                       & 20,65                      \\ \hline
\multicolumn{2}{|c|}{\textbf{SERV}}                                                                                            & 7199,93                          & 4397,00                     & 31658072                      & 86,53                      \\ \hline
\multicolumn{2}{|c|}{\textbf{QERV}}                                                                                            & 2200,03                          & 1339,38                     & 2946672                       & 25,82                      \\ \hline
\multicolumn{1}{|c|}{\multirow{4}{*}{\textbf{Fazy}}}                                                                  & 1-bit  & 7645,75                          & 4023,96                     & 30766179                      & 73,54                      \\ \cline{2-6} 
\multicolumn{1}{|c|}{}                                                                                                & 2-bit  & 4049,38                          & 2227,97                     & 9021893                       & 40,77                      \\ \cline{2-6} 
\multicolumn{1}{|c|}{}                                                                                                & 4-bit  & 2248,22                          & 1243,72                     & 2796146                       & 23,20                      \\ \cline{2-6} 
\multicolumn{1}{|c|}{}                                                                                                & 8-bit  & 1347,41                          & 799,01                      & 1076600                       & 14,83                      \\ \hline
\multicolumn{1}{|c|}{\multirow{6}{*}{\textbf{\begin{tabular}[c]{@{}c@{}}SAILOR\\ RV32I\,/\\ -Zkn-Zkt\end{tabular}}}} & 1-bit  & 3689,18\,/\,1187,49              & 1991,42\,/\,960,09          & 7346704\,/\,1140092           & 47,77\,/\,18,73            \\ \cline{2-6} 
\multicolumn{1}{|c|}{}                                                                                                & 2-bit  & 1908,06\,/\,607,89               & 1075,38\,/\,514,46          & 2051895\,/\,312733            & 24,98\,/\,9,77             \\ \cline{2-6} 
\multicolumn{1}{|c|}{}                                                                                                & 4-bit  & 1017,50\,/\,335,49               & 601,75\,/\,284,50           & 612280\,/\,95445              & 13,35\,5,39                \\ \cline{2-6} 
\multicolumn{1}{|c|}{}                                                                                                & 8-bit  & 572,22\,/\,234,09                & 395,06\,/\,185,82           & 226062\,/\,43499              & 7,54\,/\,3,79              \\ \cline{2-6} 
\multicolumn{1}{|c|}{}                                                                                                & 16-bit & 510,61\,/\,252,99                & 358,60\,/\,206,54           & 183105\,/\,52253              & 6,78\,/\,4,12              \\ \cline{2-6} 
\multicolumn{1}{|c|}{}                                                                                                & 32-bit & 630,82\,/\,174,53                & 383,85\,/\,133,81           & 242143\,/\,23354              & 7,98\,/\,2,84              \\ \hline
\end{tabular}%
}
\end{table}

%% file: tables/aes_256_encryption_benchmark_results.tex
\begin{table}[H]
\centering
\caption{Comparison of AES-256 Encryption}
\label{tab:aes_256_encryption_benchmark}
\resizebox{\columnwidth}{!}{%
\begin{tabular}{|cc|c|c|c|c|}
\hline
\multicolumn{2}{|c|}{\textbf{Processor}}                                                                                       & \textbf{Execution Time / $\mu$s} & \textbf{$E_{average}$ / nJ} & \textbf{Energy Delay Product} & \textbf{Area Time Product} \\ \hline
\multicolumn{2}{|c|}{\textbf{PicoRV32}}                                                                                        & 948,71                           & 758,49                      & 719591                        & 14,55                      \\ \hline
\multicolumn{2}{|c|}{\textbf{SERV}}                                                                                            & 5308,93                          & 3261,28                     & 17313884                      & 63,80                      \\ \hline
\multicolumn{2}{|c|}{\textbf{QERV}}                                                                                            & 1627,05                          & 992,83                      & 1615377                       & 19,09                      \\ \hline
\multicolumn{1}{|c|}{\multirow{4}{*}{\textbf{Fazy}}}                                                                  & 1-bit  & 5582,40                          & 2948,07                     & 16457281                      & 53,70                      \\ \cline{2-6} 
\multicolumn{1}{|c|}{}                                                                                                & 2-bit  & 2958,24                          & 1629,69                     & 4821027                       & 29,78                      \\ \cline{2-6} 
\multicolumn{1}{|c|}{}                                                                                                & 4-bit  & 1645,72                          & 913,54                      & 1503430                       & 16,98                      \\ \cline{2-6} 
\multicolumn{1}{|c|}{}                                                                                                & 8-bit  & 989,45                           & 587,34                      & 581141                        & 10,89                      \\ \hline
\multicolumn{1}{|c|}{\multirow{6}{*}{\textbf{\begin{tabular}[c]{@{}c@{}}SAILOR\\ RV32I\,/\\ -Zkn-Zkt\end{tabular}}}} & 1-bit  & 2723,07\,/\,496,62               & 1542,89\,/\,394,17          & 4201401\,/\,195751            & 35,26\,/\,7,83             \\ \cline{2-6} 
\multicolumn{1}{|c|}{}                                                                                                & 2-bit  & 1410,23\,/\,255,49               & 831,75\,/\,211,95           & 1172964\,/\,54152             & 18,46\,/\,4,11             \\ \cline{2-6} 
\multicolumn{1}{|c|}{}                                                                                                & 4-bit  & 753,81\,/\,143,31                & 464,80\,/\,120,32           & 350370\,/\,17244              & 9,89\,/\,2,30              \\ \cline{2-6} 
\multicolumn{1}{|c|}{}                                                                                                & 8-bit  & 425,60\,/\,103,99                & 302,64\,/ \,81,07           & 128805\,/\,8431               & 5,61\,/\,1,68              \\ \cline{2-6} 
\multicolumn{1}{|c|}{}                                                                                                & 16-bit & 358,51\,/\,117,87                & 259,78\,/\,95,24            & 93132\,/\,11226               & 4,76\,/\,1,92              \\ \cline{2-6} 
\multicolumn{1}{|c|}{}                                                                                                & 32-bit & 425,43\,/\,82,00                 & 268,40\,/\,60,98            & 114187\,/\,5000               & 5,38\,/\,1,34              \\ \hline
\end{tabular}%
}
\end{table}

%% file: tables/aes_256_decryption_benchmark_results.tex
\begin{table}[H]
\centering
\caption{Comparison of AES-256 Decryption}
\label{tab:aes_256_decryption_benchmark}
\resizebox{\columnwidth}{!}{%
\begin{tabular}{|cc|c|c|c|c|}
\hline
\multicolumn{2}{|c|}{\textbf{Processor}}                                                                                       & \textbf{Execution Time / $\mu$s} & \textbf{$E_{average}$ / nJ} & \textbf{Energy Delay Product} & \textbf{Area Time Product} \\ \hline
\multicolumn{2}{|c|}{\textbf{PicoRV32}}                                                                                        & 1576,89                          & 1259,62                     & 1986282                       & 24,18                      \\ \hline
\multicolumn{2}{|c|}{\textbf{SERV}}                                                                                            & 8384,65                          & 5120,51                     & 42933649                      & 100,77                     \\ \hline
\multicolumn{2}{|c|}{\textbf{QERV}}                                                                                            & 2562,93                          & 1560,31                     & 3998970                       & 30,08                      \\ \hline
\multicolumn{1}{|c|}{\multirow{4}{*}{\textbf{Fazy}}}                                                                  & 1-bit  & 8900,23                          & 4680,63                     & 41658692                      & 85,61                      \\ \cline{2-6} 
\multicolumn{1}{|c|}{}                                                                                                & 2-bit  & 4714,39                          & 2583,96                     & 12181782                      & 47,47                      \\ \cline{2-6} 
\multicolumn{1}{|c|}{}                                                                                                & 4-bit  & 2617,94                          & 1450,86                     & 3798271                       & 27,02                      \\ \cline{2-6} 
\multicolumn{1}{|c|}{}                                                                                                & 8-bit  & 1569,43                          & 930,36                      & 1460132                       & 17,27                      \\ \hline
\multicolumn{1}{|c|}{\multirow{6}{*}{\textbf{\begin{tabular}[c]{@{}c@{}}SAILOR\\ RV32I\,/\\ -Zkn-Zkt\end{tabular}}}} & 1-bit  & 4304,99\,/\,1400,38              & 2318,24\,/\,1130,67         & 9979988\,/\,1583363           & 55,74\,/\,22,08            \\ \cline{2-6} 
\multicolumn{1}{|c|}{}                                                                                                & 2-bit  & 2226,39\,/\,717,01               & 1251,90\,/\,605,95          & 2787216\,/\,434469            & 29,15\,/\,11,53            \\ \cline{2-6} 
\multicolumn{1}{|c|}{}                                                                                                & 4-bit  & 1187,09\,/\,396,19               & 700,26\,/\,334,90           & 831277\,/\,132684             & 15,58\,/\,6,37             \\ \cline{2-6} 
\multicolumn{1}{|c|}{}                                                                                                & 8-bit  & 667,44\,/\,277,51                & 459,27\,/\,219,65           & 306532\,/\,60955              & 8,80\,/\,4,49              \\ \cline{2-6} 
\multicolumn{1}{|c|}{}                                                                                                & 16-bit & 598,78\,/\,301,63                & 418,67\,/\,245,47           & 250689\,/\,74040              & 7,95\,/\,4,92              \\ \cline{2-6} 
\multicolumn{1}{|c|}{}                                                                                                & 32-bit & 734,33\,/\,208,02                & 444,12\,/\,158,84           & 326133\,/\,33043              & 9,29\,/\,3,39              \\ \hline
\end{tabular}%
}
\end{table}

%% file: tables/sha_256_benchmark_results.tex
\begin{table}[H]
\centering
\caption{Comparison of SHA-256}
\label{tab:sha_256_benchmark}
\resizebox{\columnwidth}{!}{%
\begin{tabular}{|cc|c|c|c|c|}
\hline
\multicolumn{2}{|c|}{\textbf{Processor}}                                                                                       & \textbf{Execution Time / $\mu$s} & \textbf{$E_{average}$ / nJ} & \textbf{Energy Delay Product} & \textbf{Area Time Product} \\ \hline
\multicolumn{2}{|c|}{\textbf{PicoRV32}}                                                                                        & 463,62                           & 373,63                      & 173223                        & 7,11                       \\ \hline
\multicolumn{2}{|c|}{\textbf{SERV}}                                                                                            & 2227,65                          & 1371,79                     & 3055861                       & 26,77                      \\ \hline
\multicolumn{2}{|c|}{\textbf{QERV}}                                                                                            & 686,53                           & 419,13                      & 287743                        & 8,06                       \\ \hline
\multicolumn{1}{|c|}{\multirow{4}{*}{\textbf{Fazy}}}                                                                  & 1-bit  & 2297,36                          & 1205,42                     & 2769295                       & 22,10                      \\ \cline{2-6} 
\multicolumn{1}{|c|}{}                                                                                                & 2-bit  & 1223,08                          & 675,75                      & 826498                        & 12,31                      \\ \cline{2-6} 
\multicolumn{1}{|c|}{}                                                                                                & 4-bit  & 685,79                           & 383,36                      & 262902                        & 7,08                       \\ \cline{2-6} 
\multicolumn{1}{|c|}{}                                                                                                & 8-bit  & 418,01                           & 251,47                      & 105119                        & 4,60                       \\ \hline
\multicolumn{1}{|c|}{\multirow{6}{*}{\textbf{\begin{tabular}[c]{@{}c@{}}SAILOR\\ RV32I\,/\\ -Zkn-Zkt\end{tabular}}}} & 1-bit  & 1205,12\,/\,695,80               & 647,99\,/\,544,46           & 780909\,/\,378838             & 15,60\,/\,10,97            \\ \cline{2-6} 
\multicolumn{1}{|c|}{}                                                                                                & 2-bit  & 622,64\,/\,351,05                & 350,86\,/\,295,30           & 218458\,/\,103666             & 8,15\,/\,5,64              \\ \cline{2-6} 
\multicolumn{1}{|c|}{}                                                                                                & 4-bit  & 338,76\,/\,178,75                & 199,87\,/\,156,71           & 67707\,/\,28012               & 4,44\,/\,2,87              \\ \cline{2-6} 
\multicolumn{1}{|c|}{}                                                                                                & 8-bit  & 219,47\,/\,92,75                 & 149,44\,/\,86,34            & 32797\,/\,8008                & 2,89\,/\,1,50              \\ \cline{2-6} 
\multicolumn{1}{|c|}{}                                                                                                & 16-bit & 206,31\,/\,50,05                 & 142,25\,/\,54,80            & 29348\,/\,2743                & 2,74\,/\,0,82              \\ \cline{2-6} 
\multicolumn{1}{|c|}{}                                                                                                & 32-bit & 230,21\,/\,42,32                 & 138,36\,/\,41,47            & 31851\,/\,1755                & 2,91\,/\,0,69              \\ \hline
\end{tabular}%
}
\end{table}

%% file: tables/sha_512_benchmark_results.tex
\begin{table}[H]
\centering
\caption{Comparison of SHA-512}
\label{tab:sha_512_decryption_benchmark}
\resizebox{\columnwidth}{!}{%
\begin{tabular}{|cc|c|c|c|c|}
\hline
\multicolumn{2}{|c|}{\textbf{Processor}}                                                                                       & \textbf{Execution Time / $\mu$s} & \textbf{$E_{average}$ / nJ} & \textbf{Energy Delay Product} & \textbf{Area Time Product} \\ \hline
\multicolumn{2}{|c|}{\textbf{PicoRV32}}                                                                                        & 1370,30                          & 1100,62                     & 1508186                       & 21,02                      \\ \hline
\multicolumn{2}{|c|}{\textbf{SERV}}                                                                                            & 6954,05                          & 4248,92                     & 29547234                      & 83,57                      \\ \hline
\multicolumn{2}{|c|}{\textbf{QERV}}                                                                                            & 2149,37                          & 1317,78                     & 2832394                       & 25,22                      \\ \hline
\multicolumn{1}{|c|}{\multirow{4}{*}{\textbf{Fazy}}}                                                                  & 1-bit  & 7101,57                          & 3724,77                     & 26451739                      & 68,31                      \\ \cline{2-6} 
\multicolumn{1}{|c|}{}                                                                                                & 2-bit  & 3782,01                          & 2100,91                     & 7945650                       & 38,08                      \\ \cline{2-6} 
\multicolumn{1}{|c|}{}                                                                                                & 4-bit  & 2122,24                          & 1176,99                     & 2497864                       & 21,90                      \\ \cline{2-6} 
\multicolumn{1}{|c|}{}                                                                                                & 8-bit  & 1292,82                          & 779,05                      & 1007176                       & 14,23                      \\ \hline
\multicolumn{1}{|c|}{\multirow{6}{*}{\textbf{\begin{tabular}[c]{@{}c@{}}SAILOR\\ RV32I\,/\\ -Zkn-Zkt\end{tabular}}}} & 1-bit  & 3851,65\,/\,2935,37              & 2011,33\,/\,2121,69         & 7746945\,/\,6227932           & 49,87\,/\,46,29            \\ \cline{2-6} 
\multicolumn{1}{|c|}{}                                                                                                & 2-bit  & 1982,77\,/\,1481,02              & 1087,55\,/\,1148,09         & 2156360\,/\,1700339           & 25,96\,/\,23,81            \\ \cline{2-6} 
\multicolumn{1}{|c|}{}                                                                                                & 4-bit  & 1065,93\,/\,753,92               & 614,62\,/\,610,37           & 655137\,/\,460173             & 13,99\,/\,12,11            \\ \cline{2-6} 
\multicolumn{1}{|c|}{}                                                                                                & 8-bit  & 668,80\,/\,390,52                & 447,43\,/\,337,92           & 299239\,/\,131963             & 8,81\,/\,6,31              \\ \cline{2-6} 
\multicolumn{1}{|c|}{}                                                                                                & 16-bit & 592,96\,/\,209,12                & 404,10\,/\,212,26           & 239616\,/\,44387              & 7,87\,/\,3,41              \\ \cline{2-6} 
\multicolumn{1}{|c|}{}                                                                                                & 32-bit & 643,21\,/\,167,74                & 386,63\,/\,155,38           & 248687\,/\,26063              & 8,14\,/\,2,73              \\ \hline
\end{tabular}%
}
\end{table}

%% file: tables/permutation_benchmark_results.tex
\begin{table}[H]
\centering
\caption{Comparison of Prince S-Box Calculation by Permutation Operations}
\label{tab:permutation_benchmark}
\resizebox{\columnwidth}{!}{%
\begin{tabular}{|cc|c|c|c|c|}
\hline
\multicolumn{2}{|c|}{\textbf{Processor}}                                                                                       & \textbf{Execution Time / $\mu$s} & \textbf{$E_{average}$ / nJ} & \textbf{Energy Delay Product} & \textbf{Area Time Product} \\ \hline
\multicolumn{2}{|c|}{\textbf{PicoRV32}}                                                                                        & 23,17                            & 18,55                       & 430                           & 0,36                       \\ \hline
\multicolumn{2}{|c|}{\textbf{SERV}}                                                                                            & 126,05                           & 77,23                       & 9735                          & 1,51                       \\ \hline
\multicolumn{2}{|c|}{\textbf{QERV}}                                                                                            & 38,45                            & 23,57                       & 906                           & 0,45                       \\ \hline
\multicolumn{1}{|c|}{\multirow{4}{*}{\textbf{Fazy}}}                                                                  & 1-bit  & 129,71                           & 68,85                       & 8931                          & 1,25                       \\ \cline{2-6} 
\multicolumn{1}{|c|}{}                                                                                                & 2-bit  & 68,75                            & 38,01                       & 2613                          & 0,69                       \\ \cline{2-6} 
\multicolumn{1}{|c|}{}                                                                                                & 4-bit  & 38,19                            & 21,28                       & 813                           & 0,39                       \\ \cline{2-6} 
\multicolumn{1}{|c|}{}                                                                                                & 8-bit  & 22,91                            & 13,66                       & 313                           & 0,25                       \\ \hline
\multicolumn{1}{|c|}{\multirow{6}{*}{\textbf{\begin{tabular}[c]{@{}c@{}}SAILOR\\ RV32I\,/\\ -Zkn-Zkt\end{tabular}}}} & 1-bit  & 66,96\,/\,20,28                  & 41,13\,/\,15,40             & 2754\,/\,312                  & 0,87\,/\,0,32              \\ \cline{2-6} 
\multicolumn{1}{|c|}{}                                                                                                & 2-bit  & 34,40\,/\,10,36                  & 22,07\,/\,8,31              & 759\,/\,86                    & 0,45\,/\,0,17              \\ \cline{2-6} 
\multicolumn{1}{|c|}{}                                                                                                & 4-bit  & 18,28\,/\,5,88                   & 12,24\,/\,4,75              & 224\,/\,28                    & 0,24\,/\,0,09              \\ \cline{2-6} 
\multicolumn{1}{|c|}{}                                                                                                & 8-bit  & 11,28\,/\,4,76                   & 8,73\,/\,3,90               & 98\,/\,19                     & 0,15\,/\,0,08              \\ \cline{2-6} 
\multicolumn{1}{|c|}{}                                                                                                & 16-bit & 9,60\,/\,6,12                    & 7,61\,/\,4,97               & 73\,/\,30                     & 0,13\,/\,0,10              \\ \cline{2-6} 
\multicolumn{1}{|c|}{}                                                                                                & 32-bit & 11,71\,/\,4,21                   & 8,18\,/\,3,35               & 96\,/\,14                     & 0,15\,/\,0,07              \\ \hline
\end{tabular}%
}
\end{table}

%% file: tables/coremark_benchmark_results.tex
\begin{table}[H]
\centering
\caption{Comparison of Coremark Benchmark}
\label{tab:coremark_benchmark}
\resizebox{\columnwidth}{!}{%
\begin{tabular}{|cc|c|c|c|c|}
\hline
\multicolumn{2}{|c|}{\textbf{Processor}}                                                                                       & \textbf{Coremark/s} & \textbf{$E_{average}$ / nJ} & \textbf{Energy Delay Product} & \textbf{Area Time Product} \\ \hline
\multicolumn{2}{|c|}{\textbf{PicoRV32}}                                                                                        & 1122,91             & 14277,15                    & 254287808                     & 273,15                     \\ \hline
\multicolumn{2}{|c|}{\textbf{SERV}}                                                                                            & 195,80              & 62124,24                    & 6345644079                    & 1227,57                    \\ \hline
\multicolumn{2}{|c|}{\textbf{QERV}}                                                                                            & 643,14              & 18944,57                    & 589128090                     & 364,93                     \\ \hline
\multicolumn{1}{|c|}{\multirow{4}{*}{\textbf{Fazy}}}                                                                  & 1-bit  & 191,26              & 54951,69                    & 5746314986                    & 1005,86                    \\ \cline{2-6} 
\multicolumn{1}{|c|}{}                                                                                                & 2-bit  & 361,36              & 30357,77                    & 1680208718                    & 557,25                     \\ \cline{2-6} 
\multicolumn{1}{|c|}{}                                                                                                & 4-bit  & 650,65              & 16976,98                    & 521850418                     & 317,20                     \\ \cline{2-6} 
\multicolumn{1}{|c|}{}                                                                                                & 8-bit  & 1084,93             & 10881,83                    & 200599939                     & 202,86                     \\ \hline
\multicolumn{1}{|c|}{\multirow{6}{*}{\textbf{\begin{tabular}[c]{@{}c@{}}SAILOR\\ RV32I\,/\\ -Zkn-Zkt\end{tabular}}}} & 1-bit  & 406,07\,/\,365,19   & 28857,03\,/\,38549,86       & 1421280773\,/\,2111226138     & 637,74\,/\,863,63          \\ \cline{2-6} 
\multicolumn{1}{|c|}{}                                                                                                & 2-bit  & 784,13\,/\,707,64   & 15563,77\,/\,21058,63       & 396969604\,/\,595176189       & 333,90\,/\,454,47          \\ \cline{2-6} 
\multicolumn{1}{|c|}{}                                                                                                & 4-bit  & 1467,05\,/\,1281,89 & 8729,06\,/\,12171,06        & 119001303\,/\,189891870       & 178,88\,/\,250,67          \\ \cline{2-6} 
\multicolumn{1}{|c|}{}                                                                                                & 8-bit  & 2594,66\,/\,1913,33 & 5700,95\,/\,8434,49         & 43943760\,/\,88165422         & 101,59\,/\,169,01          \\ \cline{2-6} 
\multicolumn{1}{|c|}{}                                                                                                & 16-bit & 4211,02\,/\,1952,85 & 3871,27\,/\,8361,13         & 18386358\,/\,85630157         & 63,07\,/\,166,92           \\ \cline{2-6} 
\multicolumn{1}{|c|}{}                                                                                                & 32-bit & 2614,87\,/\,2614,87 & 5239,27\,/\,6234,35         & 40072927\,/\,47683858         & 96,76\,/\,124,60           \\ \hline
\end{tabular}%
}
\end{table}

%% file: tables/dhrystone_benchmark_results.tex
\begin{table}[H]
\centering
\caption{Comparison of Dhrystone Benchmark}
\label{tab:dhrystone_benchmark}
\resizebox{\columnwidth}{!}{%
\begin{tabular}{|cc|c|c|c|c|}
\hline
\multicolumn{2}{|c|}{\textbf{Processor}}                                                                                       & \textbf{Dhrystone/s}    & \textbf{$E_{average}$ / nJ} & \textbf{Energy Delay Product} & \textbf{Area Time Product} \\ \hline
\multicolumn{2}{|c|}{\textbf{PicoRV32}}                                                                                        & 57007,21                & 6979,82                     & 61218732                      & 134,51                     \\ \hline
\multicolumn{2}{|c|}{\textbf{SERV}}                                                                                            & 8726,87                 & 35144,35                    & 2013572232                    & 688,56                     \\ \hline
\multicolumn{2}{|c|}{\textbf{QERV}}                                                                                            & 27883,95                & 11045,78                    & 198066951                     & 210,43                     \\ \hline
\multicolumn{1}{|c|}{\multirow{4}{*}{\textbf{Fazy}}}                                                                  & 1-bit  & 8863,80                 & 29970,22                    & 1690596654                    & 542,60                     \\ \cline{2-6} 
\multicolumn{1}{|c|}{}                                                                                                & 2-bit  & 16596,24                & 16723,66                    & 503838849                     & 303,33                     \\ \cline{2-6} 
\multicolumn{1}{|c|}{}                                                                                                & 4-bit  & 29435,42                & 9347,58                     & 158781221                     & 175,29                     \\ \cline{2-6} 
\multicolumn{1}{|c|}{}                                                                                                & 8-bit  & 48003,72                & 6222,43                     & 64812009                      & 114,62                     \\ \hline
\multicolumn{1}{|c|}{\multirow{6}{*}{\textbf{\begin{tabular}[c]{@{}c@{}}SAILOR\\ RV32I\,/\\ -Zkn-Zkt\end{tabular}}}} & 1-bit  & 15394,63\,/\,15235,14   & 19263,21\,/\,24558,35       & 625647161\,/\,805977147       & 420,55\,/\,517,53          \\ \cline{2-6} 
\multicolumn{1}{|c|}{}                                                                                                & 2-bit  & 29838,38\,/\,29538,71   & 10354,11\,/\,13450,15       & 173503254\,/\,227669823       & 219,37\,/\,272,19          \\ \cline{2-6} 
\multicolumn{1}{|c|}{}                                                                                                & 4-bit  & 56205,16\,/\,55395,65   & 5721,89\,/\,7488,86         & 50901858\,/\,67594261         & 116,72\,/\,145,02          \\ \cline{2-6} 
\multicolumn{1}{|c|}{}                                                                                                & 8-bit  & 100694,79\,/\,96796,05  & 3675,46\,/\,4494,50         & 18250512\,/\,23216348         & 65,44\,/\,83,52            \\ \cline{2-6} 
\multicolumn{1}{|c|}{}                                                                                                & 16-bit & 158969,37\,/\,146401,74 & 2519,98\,/\,3256,45         & 7926000\,/\,11121625          & 41,76\,/\,55,66            \\ \cline{2-6} 
\multicolumn{1}{|c|}{}                                                                                                & 32-bit & 184825,80\,/\,184825,80 & 2055,18\,/\,2471,52         & 5559771\,/\,6686070           & 34,22\,/\,44,07            \\ \hline
\end{tabular}%
}
\end{table}